\documentclass[9pt,twocolumn,twoside]{pnas-new}

\usepackage{xcolor}

\templatetype{pnasresearcharticle} 

\title{Tregs self-organize into a "computing ecosystem" and implement a sophisticated optimization algorithm for mediating immune response}

\author[a,1]{Robert Marsland III} 
\author[a]{Owen Howell}
\author[c]{Andreas Mayer}
\author[a,b,1]{Pankaj Mehta}

\affil[a]{Department of Physics, Boston University, Boston, MA 02215}
\affil[b]{College of Data Science, Boston University, Boston, MA 02215}
\affil[c]{Lewis-Sigler Institute, Princeton University, Princeton, NJ 08540}

\leadauthor{Marsland} 

\significancestatement{Understanding the origins and mechanisms underlying autoimmune diseases represents a major conceptual problem in immunology. It is now clear that regulatory T cells (Tregs) play an important role in suppressing autoimmune responses, but we lack an algorithmic understanding of how they function. Here we present a simple mathematical model that organizes known biology and suggests that Tregs implement a sophisticated optimization algorithm to prevent autoimmune responses. We propose novel experimental tests of these ideas and discuss implications of our results for autoimmune diseases.}

\authorcontributions{RM, AM and PM conceptualized the problem, RM, OH and PM performed analytic calculation, RM and OH performed simulations, RM and PM analyzed the results and wrote the manuscript.}
\authordeclaration{The authors declare no competing interest.}
\correspondingauthor{\textsuperscript{1}To whom correspondence should be addressed. marsland@bu.edu, pankajm@bu.edu}


\begin{abstract}
Regulatory T cells  (Tregs) play a crucial role in mediating immune response. Yet an algorithmic understanding of the role of Tregs in adaptive immunity remains lacking.  Here, we present a biophysically realistic model of Treg mediated self-tolerance in which Tregs bind to self-antigens and locally inhibit the proliferation of nearby activated T cells. By exploiting a duality between ecological dynamics and constrained optimization, we show that Tregs tile the potential antigen space while simultaneously minimizing the overlap between Treg activation profiles. We find that for sufficiently high Treg diversity, Treg mediated self-tolerance is robust to fluctuations in self-antigen concentrations but lowering the Treg diversity results in a sharp transition -- related to the Gardner transition in perceptrons -- to a regime where changes in self-antigen concentrations can result in an auto-immune response. We propose a novel experimental test of this transition in immune-deficient mice and discuss potential implications for autoimmune diseases.
\end{abstract}

\dates{This manuscript was compiled on \today}
\doi{\url{www.pnas.org/cgi/doi/10.1073/pnas.XXXXXXXXXX}}

\begin{document}

\maketitle
\thispagestyle{firststyle}
\ifthenelse{\boolean{shortarticle}}{\ifthenelse{\boolean{singlecolumn}}{\abscontentformatted}{\abscontent}}{}

\dropcap{T}he adaptive immune system of humans and other mammals solves a challenging computational problem with amazing reliability. Using only the information contained in the binding affinities between certain macromolecules, the system must distinguish potentially pathogenic cells from its own cells, in order to eliminate the former without harming the latter. While understanding how the immune system accomplishes this feat is fascinating even from a purely theoretical point of view, this problem also has many urgent practical implications since an increasing number of autoimmune diseases and allergies are thought to stem from an inability to accurately distinguish self and foreign antigens \cite{wang2015human}. 

An important step forward in the effort to understand adaptive immunity came with the discovery of regulatory T cells (Tregs)  \cite{sakaguchi1985organ,sakaguchi2008regulatory}. Like all T cell phenotypes, Tregs express T cell receptors (TCR's) on their surface, which bind to antigen peptides displayed on the surface of other cells via the major histocompatibility complex (MHC). But unlike in conventional T cells, TCR binding and activation in Tregs have the effect of suppressing T cell proliferation and cytokine production. This Treg-mediated suppression of self-activation complements negative selection in the thymus against self-reactive T cell lineages. In fact, it has been experimentally shown that even after undergoing negative selection in the thymus, T cells can raise a full immune response against native tissues if Tregs are artificially removed from the immune system \cite{sakaguchi1985organ}. For all these reasons, Tregs are thought to play a critical role preventing autoimmune responses.

The problem of distinguishing self from non-self is made even harder by the requirement that the immune system must be able to reliably respond to even small amounts of foreign antigen but be robust to potentially large fluctuations in the concentrations of self-antigens. Specifically, even in the absence of foreign antigens, the immune system needs to tolerate fluctuations in the relative abundances of self-antigens, as development, circadian cycles and other natural variations in biological activities lead to very different protein expression patterns even in healthy cells.

 In this paper, we propose a simple dynamical model for the interactions between conventional T cells, Tregs and antigens, which captures the essential aspects of known Treg biology. We then use this minimal model to understand how the immune system can achieve the stringent requirements described above, while simultaneously providing an algorithmic interpretation for Treg-mediated adaptive immunity. This work builds on a longstanding tradition of ecological modeling in theoretical immunology \cite{de1994t,levin1997mathematical,mayer2015well}, and to our knowledge is the first application of this approach to Treg dynamics. Finally, we note that for notational brevity,  in this manuscript we will always use T cells to mean conventional T cells (non-regulatory T cells).
 
\begin{figure}[t]
	\includegraphics[width=0.5\textwidth]{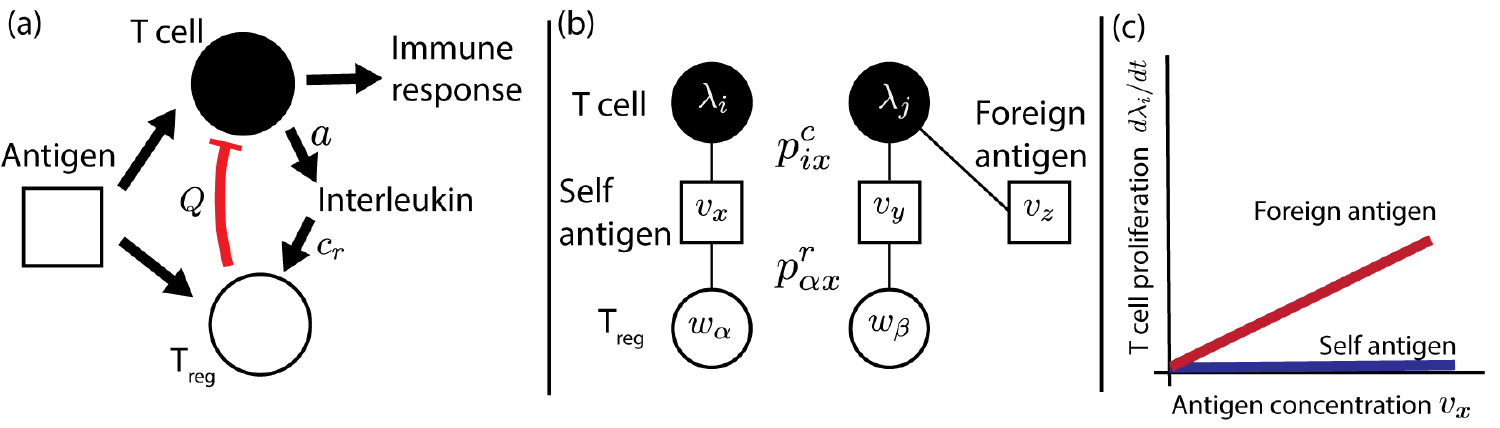}
	\caption{{\bf A minimal model of Treg-mediated self-tolerance.} \emph{(a)} Binding to antigen stimulates proliferation of conventional T cells (``T cell''), which in turn initiates an immune response. Antigen binding also stimulates proliferation of Tregs, which locally inhibit T cell proliferation, with inhibition strength $Q$. Treg proliferation also requires a sufficient local concentration of interleukin, which is produced at rate $a$ by activated T cells. \emph{(b)} Idealized ``whitelist'' network of T cell and Treg cross-reactivity functions. Each edge represents the interaction of T cell $i$ ($p_{ix}^c$) or Treg $\alpha$ ($p_{\alpha x}^r$) with antigen $x$. In this example, each kind of Treg only binds to one kind of self-antigen, and no Tregs bind to the foreign antigen. We model the growth rate of the population $\lambda_i$ of T cells of type $i$ as a function of the cross-reactivities, the abundance $v_x$ of each antigen and the abundances $w_\alpha$ of the Tregs. \emph{(c)} The ideal network ensures that T cell proliferation rates are insensitive to self-antigen concentrations $v_x$, and remain zero as levels of various proteins naturally fluctuate. At the same time, foreign antigens that do not bind to the Tregs can cause net proliferation of T cells and produce an immune response.}
	\label{fig:schem}
\end{figure}

\section*{Model Development}

We now present a minimal model for Treg-mediated self-tolerance. As shown in Figure \ref{fig:schem}, our model has three basic components: conventional T cells, Tregs, and antigens. Upon binding an antigen, T cells enter an active state in which they begin to proliferate and to release various interleukin signaling molecules. These interleukins stimulate other immune cells, including Tregs which in turn suppress T cell proliferation (Figure \ref{fig:schem}(\emph{a})). 
It is known that the activation of a sufficiently large number of T cells triggers a immune response, a process likely mediated through a quorum sensing mechanism \cite{butler2013quorum}. We do not explicitly include this last process in the minimal model presented in the main text.  Instead, we focus on the initial phase of the response and ask whether T cells will begin to proliferate in response to a foreign ligand or a change in the concentration of self ligands. We assume that sufficient proliferation will result in immune response through the processes downstream of the signaling pathways we study here. In SI Appendix (Section I), we show that the minimal model presented in the main text can be  derived from a more biologically realistic mechanistic model that includes additional components.

A typical human immune system {\color{black}has been estimated to contain} about $N_c \sim 10^6$ distinct lineages of conventional T cells and a similar number $N_r$ of Treg lineages, each carrying a different TCR, specific to a different set of antigens \cite{fazilleau2007cutting}. Labeling each T cell lineage by $i$ ($i = 1, 2 \dots N_c$) and each Treg lineage by $\alpha$ ($\alpha = 1, 2, \dots N_r$), we can encode this diversity in the cross-reactivity functions $p_{ix}^c$ and $p_{\alpha x}^r$, which quantify the strength of interaction of conventional T cells and Tregs  with  possible antigens $x$, respectively.  At the most basic level, the index $x$ simply represents a unique amino acid sequence that could be displayed on the cell surface. 
But since T cells are known to respond to antigens in a tissue-specific manner \cite{campbell2003chemokines}, $x$ can more abstractly be thought of indexing possible tissue-antigen pairs {\color{black}(see SI for more details)}. For brevity, we will often refer to these antigen-tissue pairs by the shorthand antigen. 
One can visualize the cross-reactivity functions as an interaction network, with nodes corresponding to T cells, Tregs, and antigens, and edges representing the interaction strengths (see Figure \ref{fig:schem}(\emph{b}) for a particularly simple example).

Our aim is to use these cross-reactivity functions to model the dynamics of the number of cells  $\lambda_i$ of conventional T cell lineage $i$ and the number of cells $w_\alpha$ of Treg lineage $\alpha$. In general, these abundances will depend on antigen concentrations. We will use $v_x$ to denote the abundance of antigen $x$. We will assume that the time scales on which self-antigen concentrations change is much slower than the Treg/T cell dynamics, so these $v_x$ will be treated as fixed quantities when we analyze T cell and Treg dynamics or find possible steady-states of T cell and Treg abundances. It turns out that this assumption is not essential to our main results, as shown in Supplementary Figure S3, because the independence of growth rates from $v_x$ in the emergent tiling phase implies that the same solution exists regardless of the the variations in antigen levels. 

In our minimal dynamical model, T cells of lineage $i$ can be activated at a rate proportional to the cross-reactivity functions $p_{ix}^c$  times the antigen concentration $v_x$. When activated, T cells proliferate at a rate $\rho$.  As shown in Figure \ref{fig:schem}(\emph{b}), T cell proliferation is suppressed by Tregs. Experiments indicate that Treg-mediated suppression of T cell proliferation is highly localized \cite{sakaguchi1995immunologic,takahashi1998immunologic}. This is incorporated in our model by an antigen specific suppression level $Q_x$ that is proportional to the abundance of Tregs activated by antigen $x$, with a constant of proportionality $b$. With these assumptions, conventional T cell abundances can be described using the differential equation
\begin{align}
    \frac{d\lambda_i}{dt} &= \lambda_i  \sum_x p_{ix}^c v_x (\rho-Q_x) \nonumber \\
      Q_x &= b\sum_\alpha p_{\alpha x}^r w_\alpha,
    \label{eq:Tc}
\end{align}
where the first term $p_{ix}^c \lambda_i v_x$ gives the abundance of T cells activated by each tissue-specific antigen $x$, and the second term $(\rho-Q_x)$ is the growth rate of activated cells.

Experiments also indicate that T cell activation stimulates proliferation of nearby Tregs \cite{sakaguchi2008regulatory}. One potential mechanism for this interaction is the local production of interleukin signals by activated T cells. Tregs are known to be particularly sensitive to interleukin levels and to rapidly take up interleukin from their environment \cite{malek2008biology}. We denote the local interleukin concentration  in the vicinity of cells displaying a particular  tissue-specific antigen concentration $x$  by IL$_x$. The change in number of Tregs from lineage $\alpha$, $w_\alpha$, can be written as the product of the abundance of Tregs bound to a tissue-specific antigen $x$ (given by $p_{\alpha x}^r v_x w_\alpha$)  and an interleukin-dependent local proliferation rate $c_r \mathrm{IL}_{x}$ (with  proportionality constant $c_r$). We also assume that the in absence of interleukin Tregs die at a rate $m$. These Treg dynamics can be summarized in the differential equation
\begin{align}
    \frac{dw_\alpha}{dt} &= w_\alpha \sum_x p_{\alpha x}^r v_x [c_r {\rm IL}_x-m] \nonumber \\
    {\rm IL}_x &= \frac{a \sum_j p_{jx}^c \lambda_j}{c_r \sum_\beta p_{\beta x}^r w_\beta}. 
\label{eq:Treg}
\end{align}
In the SI Appendix (Section I) we show that Eq. \ref{eq:Tc} and Eq. \ref{eq:Treg} can be derived from a realistic mechanistic model in the limit where interleukin dynamics are assumed to be fast compared to T cell and Treg proliferation. 

Surprisingly, we can rewrite these dynamics in a slightly different way that makes no explicit reference to antigens. The central objects in this formulation are the ``overlap kernels'' $\phi_{i \alpha}$ -- which measure the similarity between the activation profiles of a T cell from lineage $i$ and a Treg from lineage $\alpha$ -- and $\phi_{\alpha \beta}$ -- which measures the overlap between Tregs from lineages $\alpha$ and $\beta$ -- with
\begin{align}
\phi_{i \alpha} &= \sum_x v_x p_{\alpha x}^r p_{i x}^c \nonumber \\
\phi_{\alpha \beta} &= \sum_x v_x p_{\alpha x}^rp_{\beta x}^r. 
\label{eq:kernel_defs}
\end{align}
Notice that these overlaps depend on the cross-reactivity functions as well as the antigen concentrations.  In SI Appendix (, we describe an approximation that is exact in the ``emergent tiling'' regime discussed below, which allows the dynamics of Eq. \ref{eq:Tc} and Eq. \ref{eq:Treg} to be written entirely in terms of the overlaps:
\begin{align}
\frac{d\lambda_i}{dt} &= \lambda_i r_i \left[\rho  -b r_i^{-1} \sum_{\alpha} \phi_{i\alpha} w_\alpha \right] \nonumber \\
\frac{dw_\alpha}{dt} &= \frac{m}{\sum_\beta w_\beta \bar{p}_\beta} w_\alpha \left[ \kappa_\alpha - \sum_\beta  \phi_{\alpha\beta} w_\beta\right],
\label{eq:niche-dynamics}
\end{align}
with $\kappa_\alpha= a \sum_j \lambda_j \phi_{j \alpha}/m$, $r_i = \sum_x v_x p_{ix}$ and $\bar{p}_\beta = \sum_x p_{\beta x}^r$. Surprisingly the explicit dependence on antigens has completely disappeared from Eq. \ref{eq:niche-dynamics}. Instead, all information about antigen concentrations appears only through the overlap kernels. This is similar to the ``kernel  trick'' in Machine Learning where all information about overlaps in a feature space can be encoded in a kernel function \cite{bishop2006pattern}.

These new equations also naturally lend themselves to an ecological interpretation in terms of Consumer Resource Models and generalized Lotka-Volterra models. We can view T cells as exponentially growing ``resources'', with growth rate $r_i^{-1} \rho$, that are consumed by Tregs, with the consumption rate depending on the ``resource utilization function'' $\phi_{i \alpha}$. Notice that the Treg dynamics in Eq. \ref{eq:niche-dynamics} take the form of a Lotka-Volterra equation. Tregs grow at a rate $\kappa_\alpha$ that depends how many resources they consume but also compete with other Tregs.  The strength of competition depends on the overlap kernel $\phi_{\alpha \beta}$. Ecologically, $\phi_{\alpha \beta}$  can be thought of as  the``niche-overlap'' between Tregs in antigen space. The mechanistic origins of this competition can be traced to the fact that higher niche-overlaps mean Tregs are more likely to be colocalize and hence more likely to compete for interleukins produced by T cells at a given tissue/antigen (see Fig. \ref{fig:schem}).

\section*{Results}

We now analyze the implications of these dynamics in greater detail. Our analysis exploits the ecological interpretation described above by making use of our recently discovered mapping between ecological dynamics and constrained optimization \cite{mehta2019constrained,howell2020machine,marsland2020minimum}. This allows us to naturally give an algorithmic interpretation of the computations performed by Tregs and identify a novel phase transition in the behavior of Treg mediated self-tolerance as a function of Treg diversity.

\begin{figure}[t]
	\includegraphics[width=0.5\textwidth]{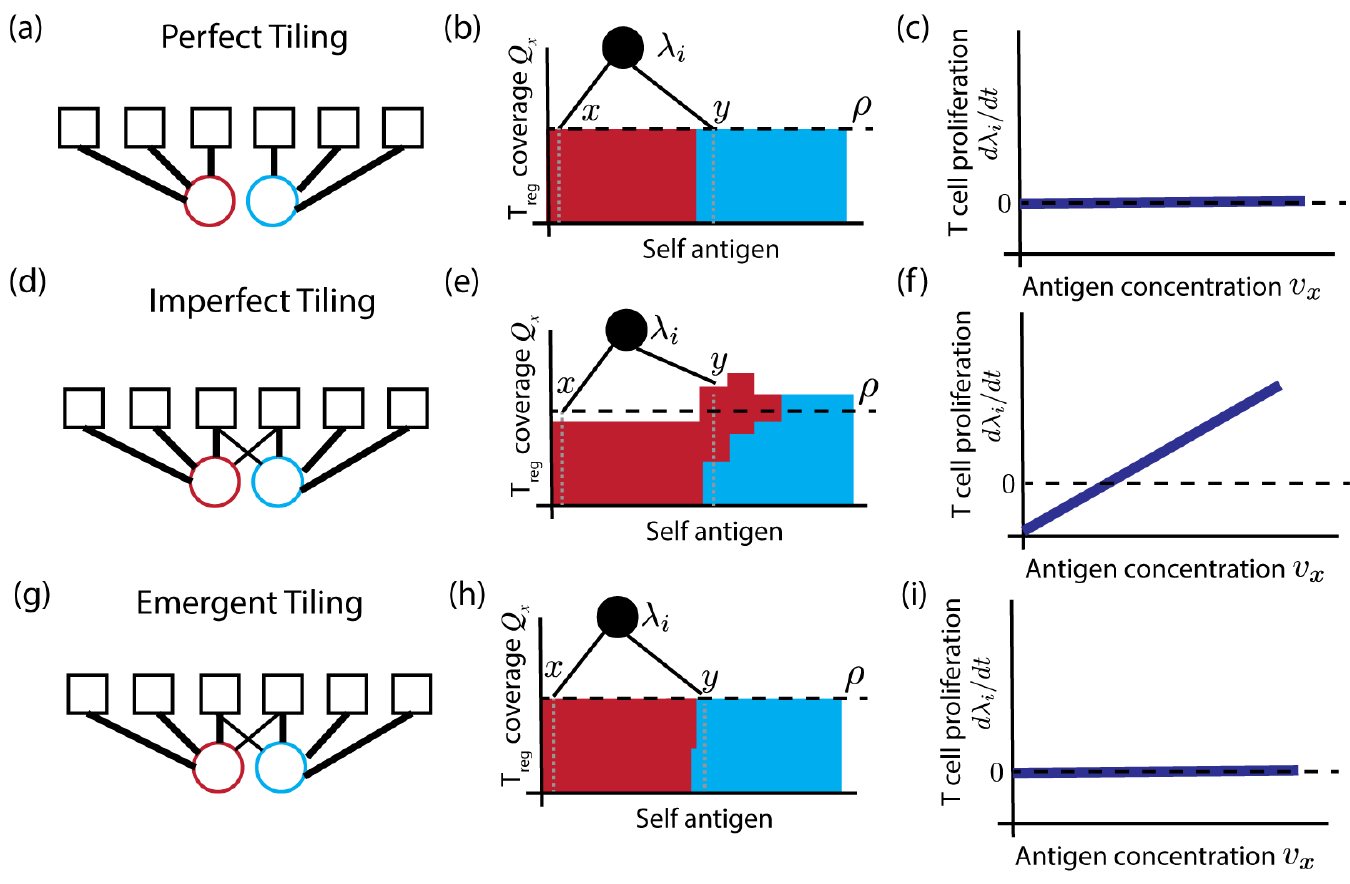}
	\caption{{\bf Emergent tiling of self-antigen space required for robust self-tolerance.} \emph{(a)} Cross-reactivity network with perfect tiling of self-antigens by Tregs. Each self-antigen binds to exactly one Treg receptor, and all nonzero affinities are equal. \emph{(b)} Local Treg-mediated suppression in this scenario, with the height of the each colored region representing the suppression level contributed by a given Treg in the vicinity of a given self-antigen. The two Treg lineages are shown in the same colors as the first panel, and antigens are arranged along the horizontal axis. Perfect tiling ensures that the suppression strength is uniform across all self-antigens, exactly canceling the basal proliferation rate $\rho$. Also shown is a conventional T cell (black circle) that binds to two of the antigens ($x$ and $y$). \emph{(c)} Net proliferation rate of T cell from previous panel. Rate is shown as a function of the concentration $v_x$ of antigen $x$, assuming that the concentration $v_y$ of antigen $y$ is held fixed. \emph{(d)} Generic cross-reactivity network, with non-uniform affinities and overlap between the cross-reactivity functions of different Tregs. \emph{(e)} Possible result for Treg-mediated suppression, which is no longer uniform across antigens. \emph{(f)} Proliferation rate of T cell from previous panel as a function of $v_x$ at fixed $v_y$. \emph{(g)} Generic cross-reactivity network, as above. \emph{(h)} Treg coverage under ``emergent tiling,'' where a set of Treg abundances $w_\alpha$ is found that restores uniform suppression levels, despite the heterogeneity and overlaps in the cross-reactivity network. Note that this solution is generically only possible at much higher levels of Treg diversity for the given number of antigens. \emph{(i)} Proliferation rate of T cell from previous panel as a function of $v_x$ at fixed $v_y$.}
	\label{fig:idea}
\end{figure}

\subsection*{Tregs  minimize "niche-overlap" in antigen space}

If Treg and conventional T cell dynamics are fast compared to the the rate at which antigen concentrations change, we can focus on analyzing the steady-state abundances of Tregs and T cells. In this case, we can set the left hand side of equations in Eq. \ref{eq:niche-dynamics} equal to zero and the resulting steady-state equations have a natural interpretation in terms of constrained optimization. In SI Appendix (Section II), we show that the steady states of the dynamical equations Eq. \ref{eq:niche-dynamics} are equivalent to the solutions of the following the constrained optimization problem:
\begin{align}
\underset{\mathbf{w}}{\rm argmin} \, &\frac{1}{2}\sum_{\alpha,\beta} w_\alpha \phi_{\alpha\beta} w_\beta \nonumber \\
{\rm subject}\,{\rm to:}& \,\,\, r_i^{-1} \sum_\alpha \phi_{i\alpha} w_\alpha \geq \frac{\rho}{b} \,\, {\rm and} \,\, w_{\alpha} \geq 0.
\label{Eq:w_opt}
\end{align}
The T cell concentrations $\lambda_i$ play the role of generalized Lagrange multipliers in the Karush-Kuhn-Tucker (KKT) conditions that enforce the inequality constraints above.

This optimization problem has a very beautiful biological interpretation. To see this, note that the $\phi_{\alpha\beta}$ is a measure of the similarity between Tregs. Thus, this optimization tells us that Treg populations self organize to minimize overlaps in the activation profile of Tregs. However, when performing this optimization, one must ensure that no T cell lineage $i$ can be activated in the absence of foreign ligands to prevent undesirable autoimmune responses. This last condition is represented in the inequality constraint $r_i^{-1} \sum_\alpha \phi_{i\alpha} w_\alpha \geq \frac{\rho}{b}$ which simply states that Treg must be able to suppress the proliferation of any T cell lineage $i$. Equivalently, from an ecological perspective, this means that Tregs must be able to cover the potential space of activated T cells while simultaneously minimizing niche overlap between Tregs.  This is analogous to niche partitioning and species packing in ecological systems \cite{MacArthur1969,MacArthur1970}, with Treg lineages playing the roles of species and T cells playing the role of resources.

\subsection*{Emergent tiling is required for robust self-tolerance}

The previous results relied on the overlap kernel formulation of our model. We now reanalyze these dynamics from the perspective of antigens.
As discussed in the introduction, an important property of an effective immune system is that it must reliably respond to foreign antigens but also be robust to fluctuations in the concentrations of self-antigens. One simple, but biophysically unrealistic, way of achieving this robustness is depicted in Figure \ref{fig:schem}\emph{(b)}. The figure depicts a simple ``whitelist'' scenario for Treg-mediated self-tolerance, with a specialized Treg for each self-antigen and all Tregs having the same binding affinity. In such a scenario, the sums over $x$ in Eqs. (\ref{eq:Tc}-\ref{eq:Treg}) are unnecessary, since each equation contains only one term. Setting $d\lambda_i/dt = dw_\alpha/dt = 0$ then yields uniform equilibrium Treg coverage $Q_x = \rho$ and a uniform interleukin profile ${\rm IL}_x = m/c_r$ over all antigens $x$ that interact with a surviving T cell and Treg. As can be immediately seen from the equations, these conditions guarantee that proliferation rates $d\lambda_i/dt$ and $dw_\alpha/dt$ remain zero for all choices of the antigen concentrations $v_x$. A direct consequence of this is that Treg-mediated self tolerance is robust against fluctuations in the antigen abundances. Nonetheless, if a foreign antigen is introduced that does not interact with any of the Tregs, T cells that bind to the new antigen will still proliferate at the original rate $\rho$ since there is no specialized Treg to inhibit their growth.

Real immune systems, however, cannot achieve this one-to-one correspondence between self-antigens and Tregs because of biophysical constraints stemming from the nature of TCR-peptide interactions \cite{sewell2012must}. Figure \ref{fig:idea} illustrates three other possible scenarios of response to fluctuations in self-antigen concentrations (Fig. \ref{fig:idea}(\emph{a}-\emph{c})). In the first scenario, each Treg binds with equal strength to a different set of non-overlapping antigens. These sets perfectly tile the set of self-antigens, covering all of them without gaps or overlaps. Since each antigen interacts with only one Treg (though Tregs will generically interact with multiple antigens), a set of Treg abundances can still easily be found for which the proliferation rate vanishes everywhere.  However, just as in the original whitelist example, this scenario can only achieved by requiring a biophysically implausible fine tuning of TCR-peptide binding. In the absence of such fine tuning, the system becomes sensitive to fluctuations in self-antigen abundances $v_x$ since generically the Treg coverage of antigen space will be  uneven  (Fig. \ref{fig:idea}(\emph{d}-\emph{f})).

These examples suggest that the only biophysically plausible way for the immune system to achieve robustness to fluctuations in self-antigens while maintaining sensitivity to foreign antigens is to tune the relative abundances of the Tregs and T cells to produce uniform coverage and a uniform interleukin profile, with $Q_x = \rho$ and ${\rm IL}_x = m/c_r$ at every antigen $x$, despite the biophysically unavoidable overlaps between the Treg cross-reactivity functions (Fig. \ref{fig:idea}(\emph{d}-\emph{f})). We call such a tilling of the antigen space an ``emergent tiling.'' Note that emergent tiling is generically impossible in the simple cross-reactivity network sketched here, with many antigens and only two Tregs, and the schematic is only meant to convey the idea of uniform coverage in the presence of overlaps. We now proceed to investigate the conditions for emergent tiling without fine-tuning in more complex networks.

\begin{figure}[t]
	\includegraphics[width=0.5\textwidth]{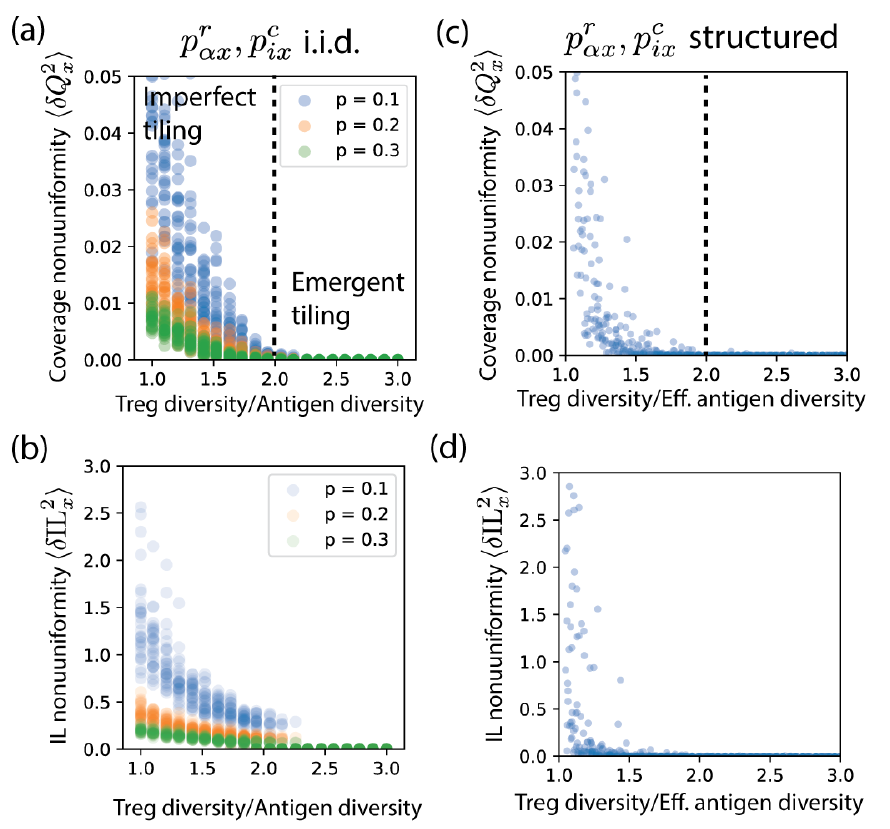}
	\caption{{\bf A phase transition from imperfect tiling to emergent tiling as a function of Treg diversity.} \emph{(a)} Mean squared deviation $\langle \delta Q_x^2\rangle$ of Treg coverage from the ideal uniform solution $Q_x = \rho$ versus the ratio $N_r/N_a$ of the number of Treg lineages to the number of antigens. Equilibrium Treg and T cell abundances were obtained by convex optimization as described in the main text and SI. $N_a$ and the number of conventional T cell lineages $N_c$ were held fixed at 100 and 1,000, respectively, while $N_r$ was swept from 100 to 300. The elements of the cross-reactivity functions $p_{\alpha x}^r$ and $p_{ix}^c$ were sampled independently from Bernoulli distributions with success probability $p$ from 0.1 to 0.3. \emph{(b)} Mean squared deviation $\langle \delta {\rm IL}_x^2\rangle$ of local interleukin levels from the ideal uniform solution ${\rm IL}_x = m/c_r$, for the same set of simulations. \emph{(c)} Same as first panel, but with structured cross-reactivity functions $p_{\alpha x}^r$ and $p_{ix}^c$ encoding a one-dimensional shape space, as described in the main text and SI. The horizontal axis indicates the ratio of the effective number of Tregs $N_r^{\rm eff}$ to $N_a$, where $N_r^{\rm eff}$ is defined as the number of singular values of the cross-reactivity matrix $p_{\alpha x}^r$ that exceed a cutoff threshold of $\epsilon = 10^{-6}$. \emph{(d)} Mean squared deviation of local interleukin levels for the same set of simulations.}
	\label{fig:d}
\end{figure}

\subsection*{Emergent tiling is possible above a threshold level of Treg diversity}

Inserting the full expressions for ${\rm IL}_x$ and $Q_x$ from Eqs. \ref{eq:Tc}-\ref{eq:Treg} into the conditions for uniform coverage discussed above ($Q_x = \rho$, ${\rm IL}_x = m/c_r$), we find that sufficient conditions for a stable emergent tiling solution can be written as
 \begin{align}
\sum_{\alpha} p_{\alpha x}^r w_\alpha = \frac{\rho}{b}  \hspace{0.5cm} {\rm and} \hspace{0.5cm} \sum_{j} p_{j x}^c \lambda_j  = {m \rho \over ab},
 \label{eq:ET}
 \end{align}
where as before $\alpha$ runs over Treg lineages, $j$ over T cell lineages, and $x$ over possible self antigens. Note that we have used the first condition to simplify the expression for ${\rm IL}_x$ in the second. When a solution to these equations exists, it also solves the optimization problem stated in Eqs. \ref{Eq:w_opt} above, but this is not immediately clear from the formulation in terms of overlaps. In SI Appendix (Section II), we perform a series of duality transformations to rewrite the optimization problem in such a way that Eqs. \ref{eq:ET} naturally arise.
In general, whether these conditions can be satisfied will depend on the dimensionality of the space we are working in (i.e. the number of Tregs $N_r$, the number of T cells $N_c$,  the number of tissue specific antigens $N_a$, and the structure of the cross-reactivity function).
 
 The question of whether the Treg dynamics achieve emergent tiling is thus reduced to the question of whether the equations in Eq. \ref{eq:ET} have solutions. We begin by investigating the simplest kind of cross-reactivity function, where the entries of the matrix $p_{\alpha x}^r$ and $p_{i x}^c$ are independently drawn from a Bernoulli distribution. 
( {\color{black} As we discuss below, real cross-reactivity functions are much more complicated than this but it is helpful theoretically to start with this simple model to gain intuition.} This problem has been considered in various contexts including the theory of perceptrons (a simple model of statistical learning) and more recently in the context of ecology \cite{gardner1988space,Nishimori,landmann2018systems}. In fact, it is possible to show that in high dimensions when $N_r, N_c, N_a \gg 1$  a solution generically exists if there are at least twice as many T cell and Treg lineages as types of antigens:  $N_r/N_a>2$ and $N_c/N_a >2$ and no solution exists if the opposite is true. These two regimes are separated by a phase transition known as the Gardner transition in the statistical physics literature \cite{gardner1988space, gardner1987maximum, landmann2018systems}. We give a brief overview of this relation in SI Appendix (Section IIID).
 
In Figure \ref{fig:d}\emph{(a)-(b)}, we plot the equilibrium deviations from uniform Treg coverage $\langle \delta Q_x^2\rangle = \frac{1}{N_a}\sum_x (Q_x - \rho)^2$ and from uniform interleukin concentration $\langle \delta {\rm IL}_x^2\rangle = \frac{1}{N_a}\sum_x ({\rm IL}_x - m/c_r)^2$ as a function of the ratio $N_r/N_a$, for randomly generated cross-reactivity functions with $N_a = 100$ antigens, with $p_{\alpha x}^r$ sampled from Bernoulli distributions with three different probabilities of interaction. To make these plots, we solved the corresponding constrained optimization problem (Eq. \ref{Eq:w_opt}) using standard numerical techniques for convex optimization (see SI Appendix Section IV). We see that in all cases, the nonuniformity vanishes near the predicted transition point $N_r/N_a = 2$.

To determine the number of Treg lineages $N_r$ needed to achieve emergent tiling, we must estimate the number of distinct self-antigens $N_a$. The maximum number of cleavage points for creating a peptide for display on the MHC is the total number of codons ($\sim  10^{7}$) in the coding regions of the human genome. Since {\color{black}some} of these peptides are redundant, and only a fraction can be successfully cleaved and loaded onto MHC's, a conservative upper bound would have 10 percent of the possible cleavage sites result in displayed peptides, yielding $N_a \sim 10^6$. This number is the same order of magnitude as the observed Treg diversity, which has been estimated at $3.5\times 10^6$ \cite{fazilleau2007cutting}. It is therefore not  implausible that $N_r/N_a>2$ in real immune systems.

However, biologically realistic cross-reactivity functions differ {\color{black} significantly} from the i.i.d cases described above since antigens with similar shapes bind to similar sets of receptors. {\color{black} Receptor affinities in TRegs and conventional T cells are shaped through a complex process of positive and negative selection in the 
Thymus. For example, experiments suggest that Tregs may in fact have higher affinity for self antigens than conventional T cells \cite{hsieh2012selection}. In general, quantitatively understanding cross-reactivity functions in different cell types is an important open problem. In light of this incredible and unknown complexity, we sought to ask if the intuition above also holds for slightly more biologically realistic scenarios. To do so, we used a toy-models for cross-reactivity based on antigen low-dimensional shape spaces that have been extensively used in the statistical physics literature to model TCR repertoires \cite{altan2020quantitative}.} Fig. \ref{fig:d}\emph{(c)-(d)} shows numerical simulations of a one-dimensional shape space with $N_a = 5,000$ antigens. Each of $N_r = 500$ Tregs and $N_c = 500$ conventional T cells binds to a group of similar antigens, with $p_{\alpha x}^r = e^{-(x-x_\alpha)^2/2\sigma^2},\, p_{i x}^c = e^{-(x-x_i)^2/2\sigma^2}$, where the center of the group $x_\alpha$ or $x_i$ is randomly chosen for each Treg $\alpha$ and T cell $i$, and the width $\sigma$ is the same for all lineages in a given simulation run (see SI Appendix for details).  Since the antigens are now ``correlated'', instead of the absolute number of antigens it is useful do define an effective antigen dimension $N_a^{\rm eff}$. We define $N_a^{\rm eff}$ as the effective rank of the cross-reactivity matrix $p_{\alpha x}^r$, that is, the number of singular values that exceed a cutoff threshold $\epsilon = 10^{-6}$. The effective number of antigens decreases from $N_a^{\rm eff} = 484$ to 97 as $\sigma$ varies from 10 to 100. When we plot the nonuniformity $\langle \delta Q_x^2\rangle$ and $\langle \delta {\rm IL}_x^2\rangle$ as a function of $N_r/N_a^{\rm eff}$, we see a very similar pattern to the i.i.d. case, with reliable emergent tiling for $N_r/N_a^{\rm eff} > 2$. 
{\color{black}As a further check on our model, we also ran simulations when the cross-reactivities are drawn from a five-dimensional shape space (see SI Appendix Section IV, Fig. S4 and Fig. S5). Once again, we found that an emergent tiling phase when  $N_r/N_a^{\rm eff}>2$. We also ran simulations to check what happens when we allow the  ``antigen concentrations'' $v_x$ to vary in time.  SI Appendix, Fig. S3 show simulations from our full model when  $v_x$ is allowed to rapidly oscillate in time. Somewhat surprisingly, we find that the dynamics still self-organizes into an emergent tiling phase despite the fact that Treg and Tcell abundances no-longer reach a steady-state. Collectively, these results suggest that the existence of the emergent tiling phase is a robust feature of this class of models.}{\color{black} Finally, we note that for more biologically realistic choices of the cross-reactivity matrices while the exact ratio of Treg to antigen diversity at which the transition to the emergent tiling phase occurs may change, general arguments from statistical mechanics suggest that such a phase will exist even in these more complex settings.}

\begin{figure}[t]
	\includegraphics[width=0.5\textwidth]{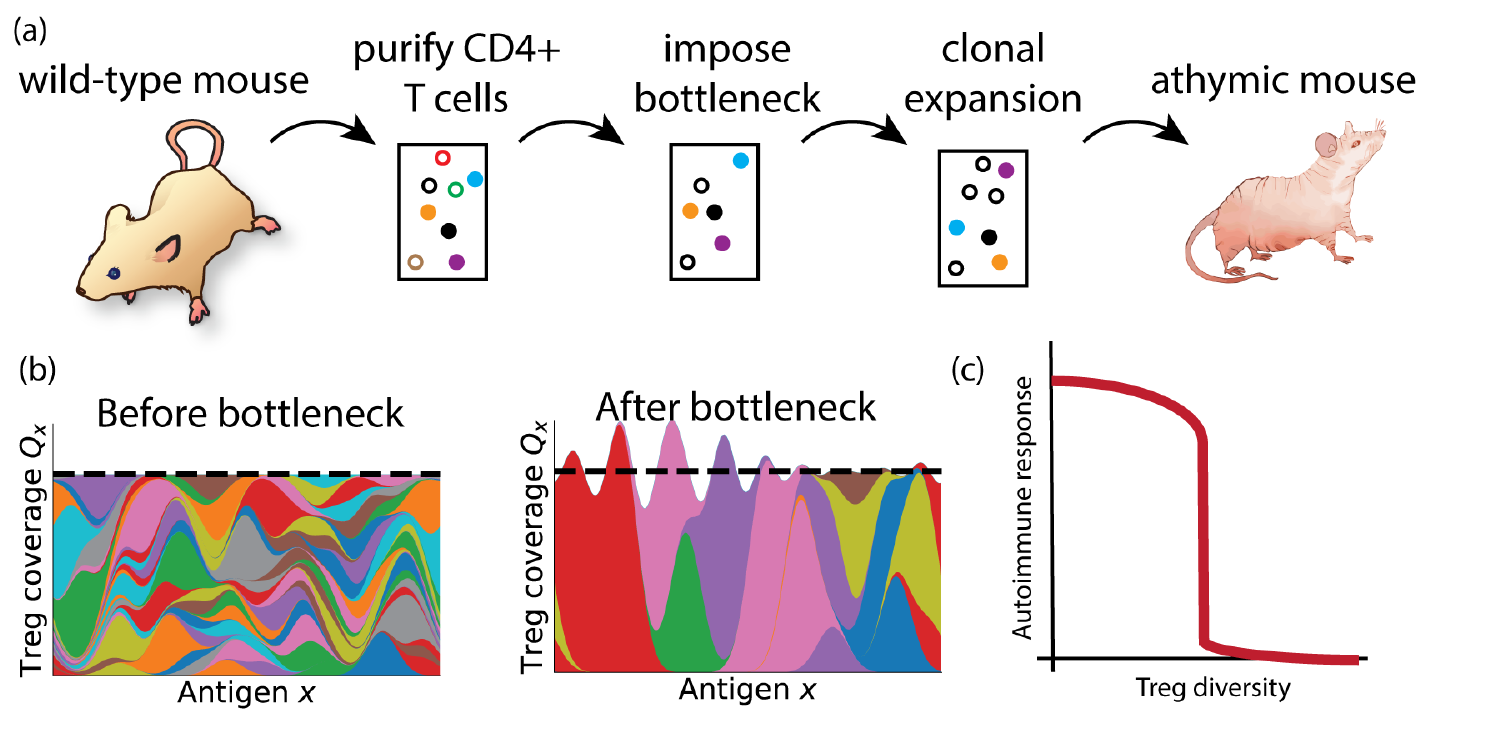}
	\caption{{\bf Proposed experimental test for emergent tiling transition.} \emph{(a)} T cells (circles) are purified from a wild-type mouse, and Tregs are selectively depleted from the sample following standard protocols (e.g., antibody plus complement). The number of Treg clonotypes (open circles) remaining after depletion will depend on the size of the resulting population bottleneck, allowing a variety of levels of Treg diversity to be generated. The total Treg cell count is then restored to its original level using standard protocols for clonal expansion of Tregs. These Tregs, along with the rest of the T cells from the original sample, are then introduced into athymic nude mice, which lack native T cells. After an appropriate waiting period, the mouse is evaluated for an autoimmune response, for instance by measuring levels of antibodies against various self-peptides. \emph{(b)} Schematic of Treg coverage before and after the population bottleneck. As in Fig. \ref{fig:idea}, each colored region represents a different Treg, and the height of the region represents the contribution of that Treg to the total suppression $Q_x$. The wild-type mouse has sufficient Tregs to achieve emergent tiling (left), but this is no longer possible if the Treg diversity is driven below the threshold (right). \emph{(c)} Schematic of predicted autoimmune activity (e.g., self-antibody levels) as a function of Treg diversity. Our model predicts a sharp transition at a critical diversity level, which is significantly higher than what is required simply to cover all the self-antigens.}
	\label{fig:path}
\end{figure}

\subsection*{Proposed experimental test of emergent tiling transition}
A key prediction of our previous analysis is that lowering the Treg diversity results in a sharp transition  to a regime where changes in self-antigen concentrations can result in an auto-immune response (see Fig. \ref{fig:d}). We now propose an experimental test of this prediction by  repurposing a classical immunological experimental design previously used to discover Treg function (see Fig. \ref{fig:path}). In the original experiments T cells (including both Tregs and conventional T cells) were transferred from the spleen and/or lymph nodes of a mouse with a functioning thymus to another mouse from a strain (``athymic nude'') homozygous for a mutation that renders it congenitally incapable of producing any kind of T cell \cite{sakaguchi1985organ,sakaguchi1995immunologic}. If all the T cells are transferred together, the recipient mouse remains healthy. But if the Tregs are eliminated from the population, with only the conventional T cells injected into the recipient, severe autoimmune syndromes result in multiple organs.

In the decades since these original studies were carried out, new techniques have been developed that make it possible to modulate and assess the impact of Treg repertoire diversity. First of all, it was shown that Tregs can undergo clonal expansion \emph{in vitro} after co-culture with dendritic cells \cite{yamazaki2003direct}. Specific Treg clones can be expanded using known receptor-antigen pairs, or all the Tregs can be expanded together, nonspecifically, by loading the dendritic cells with antibodies against CD3, one of the components of the TCR, which effectively acts like a universal antigen. Secondly, repertoire diversity can be directly measured with high-throughput sequencing of the TCR genes \cite{fazilleau2007cutting,sherwood2011deep,six2013past}.

These two techniques together allow one to envision experiments with Treg ecology analogous to existing methods in microbial ecology \cite{tecon2019bridging}. In particular, one could reduce the Treg repertoire diversity by imposing a population bottleneck. Clonal expansion could then be applied to bring the total cell count back to the original level, before injecting the cells into the recipient organism. Receptor sequencing of the post-expansion cells makes it possible to quantify the final Treg diversity, and calibrate the bottleneck to achieve a range of diversity levels. 

Our theory predicts that emergent tiling is required for robust self-tolerance, and that emergent tiling is only possible if the Treg diversity exceeds a certain threshold, which is larger than the bare minimum level required to simply cover all the self-antigens. If the diversity is reduced to sufficiently low levels, the Treg populations can only achieve imperfect tiling, and conventional T cell proliferation can be induced by natural fluctuations in self-antigen concentrations. Measures of autoimmune activity, such as concentrations of antibodies against self-peptides from various organs \cite{sakaguchi1995immunologic}, should therefore show a strong negative correlation with the number of distinct Treg clones injected into the athymic mouse, with a sharp Treg diversity threshold below which there is an autoimmune response.

\subsection*{Comparison with Existing Data}
{\color{black}
Our model also makes several qualitative predictions that can be compared with existing experimental observations. First of all, we predict that autoimmune disorders can be caused by genetic mutations that significantly restrict the Treg TCR repertoire but otherwise leave the immune system intact. This prediction has been confirmed in non-obese diabetic (NOD) mice, a standard animal model for type I diabetes \cite{ferreira2009non}. These mice spontaneously develop an autoimmune disorder whereby the immune system destroys the insulin-producing $\beta$ cells in the pancreas. An assay of thymic TCR repertoires from these mice revealed that the diversity of insertions/deletions in one subfamily of $\alpha$-chain V domains was between 5 and 8 times smaller for Tregs than for conventional T cells, while the diversity levels of the two lineages were similar (within a factor of 2) for wild-type (C57BL/6) mice. Furthermore, the NOD mice only expressed between five and eight of the 49 functional $\alpha$-chain J domains in the mouse genome, as compared to more than 20 in each of the wild-type mice, resulting in a further repertoire restriction. This data suggests that the genetic defects of the NOD mice lead to excessively stringent thymic selection criteria for commitment to the Treg lineage, resulting in a Treg repertoire size below the threshold for emergent tiling. 

The relationship between Treg diversity and self-tolerance has also been tested using transgenic mice engineered to eliminate all variation in the TCR $\beta$ chain \cite{adeegbe2010cd4}. In a similar setup to the one proposed above, a strain of mice that is congenitally deficient in Tregs (due to deletion of the Treg interleukin receptor) is injected with Tregs purified from either a wild-type mouse or one of the low-TCR-diversity transgenic mice. While the Tregs from the wild-type mouse reliably prevented autoimmune pathologies, most of the grafts from the low-diversity transgenic mice resulted in some autoimmune symptoms. The low-diversity Treg injection was sufficient, however, to keep five out of 16 mice healthy for the duration of the experiment. This suggests that the constricted Treg repertoire of the transgenic mice may have been sufficient to provide full coverage of self-antigens but too small to achieve emergent tiling. Our model predicts that diversity levels in this range can provide some short-term protection against autoimmunity, especially in a highly controlled laboratory setting, but that this protection is easily lost due to changes in relative antigen abundances $v_x$. 

In this same series of Treg transfer experiments, the authors also explored some features of the Treg population dynamics. By quantifying the frequencies of individual V$\alpha$2 CD3 sequences before injection into the host and at several time points afterwards, they showed that the relative population sizes of Treg clones changed dramatically over the course of three months \cite{adeegbe2010cd4}. Many of the dominant clones at the end of the experiment were extremely rare at the beginning, while initially dominant clones were nearly driven to extinction. This repertoire reshaping agrees with the predictions of our model, where the high interleukin levels in Treg-deficient mice lead to rapid proliferation of the exogenous Tregs, with the proliferation rates of different clones highly dependent on the distribution of displayed self-antigens within the host.

Finally, our model makes predictions about the results of injecting exogenous Tregs into an animal whose own Treg populations are already established. If the recipient has high enough Treg diversity to achieve emergent tiling, the uniform interleukin profile barely meets the minimum level required for proliferation. This means that no new Tregs will be able to proliferate, regardless of their binding specificity. If the recipient is not in the emergent tiling phase, however, some of the local interleukin levels will be sufficient to support net proliferation, and novel Tregs specific for the corresponding antigens will be able to invade. This prediction is consistent with experiments on TCR transgenic mice with Treg diversity reduced to about half the level of the wild-type \cite{fohse2011high}. When Tregs from wild-type mice were injected into these transgenic mice, they proliferated and eventually made up more than 20\% of the overall Treg population in the recipient. But when Tregs from the transgenic mice were injected into the wild-type, no proliferation was observed.

All these qualitative observations admit of many possible explanations, but together they demonstrate the power of our model for providing a single underlying biologically realistic mechanism that is parsimonious with a wide range of experimental data on Treg function. }

\section*{Discussion}
In this work, we have constructed a minimal model of Treg-mediated self-tolerance that is consistent with known biological and experimental facts. Specifically, our model reflects the empirical observation that Tregs suppress the proliferation of conventional T cells via a highly local mechanism \cite{takahashi1998immunologic,sakaguchi1995immunologic}, which depends on the binding specificity of the Treg TCR \cite{hori2002specificity,olivares2000repertoire}. This feature of the biology makes the overlap $\phi_{i\alpha}$ between binding profiles of Tregs and conventional T cells a crucial quantity in the analysis of Treg function. Our model also accounts for the dependence of Treg proliferation on local concentrations of interleukins generated  by activated conventional T cells \cite{sakaguchi2008regulatory,malek2008biology}. It has long been recognized that this stimulatory effect of conventional T cells on Tregs completes a homeostatic feedback loop (see Fig. 2 of \cite{sakaguchi2008regulatory}). But the fact that  interleukin is internalized and degraded after binding to a Treg's interleukin receptor \cite{malek2008biology} allows us to formulate a more specific hypothesis: that Treg population dynamics can be modeled using classical resource competition theory, with local interleukin pools acting as scarce limiting resources. This analogy highlights the role of the overlap $\phi_{\alpha \beta}$ between Treg binding profiles as a second crucial quantity, and enables us to analyze the Treg behavior in terms of the ecological concepts of species packing and niche partitioning \cite{MacArthur1969}. 

We have shown that the resulting immune dynamics have a natural interpretation in terms of optimization. We also find that for sufficiently high Treg diversities, this simple dynamics allows Tregs to self-organize into a state that allows the immune system to retain sensitivity to foreign ligands while simultaneously being robust to fluctuations in the concentrations of self antigens.

The high Treg diversity required for emergent tiling helps explain the otherwise surprising fact that Tregs contain a similar number of distinct TCR's as conventional T cells, even though the Tregs only need to interact with peptides from the human genome, while the conventional T cells must cover all possible pathogens \cite{fazilleau2007cutting}. Even though a much smaller number of Treg lineages would be sufficient to cover all the self-antigens, emergent tiling {\color{black} in our simple models} requires $N_r/N_a>2$, with $N_a \sim 10^6$ in the human immune system. 

This theory raises a number of further questions, which we do not address here. The first concerns the specific requirements for maintaining pathogen sensitivity. While the repertoire of conventional T cells in an ideal immune system should cover all possible foreign antigens, the repertoire of Tregs must have some gaps in coverage in order to leave the system free to respond to at least some subset of foreign antigens. It has been observed that high-affinity antigen binding is associated with commitment to the Treg phenotype, suggesting a difference in positive and/or negative selection thresholds between Tregs and other phenotypes \cite{sakaguchi2008regulatory, kovsmrlj2008thymus, kovsmrlj2009thymic}. This seems to be a natural way of achieving this difference in coverage, and understanding the details of how this might be achieved is an important area of future research. {\color{black} Doing so will require us to move beyond the simple shape space models for cross-reactivity used in this work and consider more biophysically
realistic models for antigen-TCR binding.}

Another interesting question is about the acquisition of tolerance to foreign peptides. The immune system is tolerant to many things that are not presented in the thymus when the Tregs are generated, such as peptides from various foods, and commensal microbes \cite{coombes2007functionally}. Tolerance can also be acquired later in life. There is evidence that Tregs can be generated from other T cell lineages in the periphery, and not just in the thymus \cite{coombes2007functionally,sakaguchi2008regulatory}. This raises the possibility that the Treg repertoire may be adaptively repopulated on a slow timescale with new lineages that interact with such non-self peptides. 

Our theoretical framework allows us to generate hypotheses about causes of natural autoimmune disorders. One robust observation that has yet to be explained is the de novo onset of autoimmune syndromes in people with persistent viral infections \cite{kivity2009infections}. A possible reason for this is that the characteristics of a persistent infection somehow do not allow for adaptation via generation of new Tregs that specifically bind to the viral antigen, and so existing Treg populations end up clonally expanding to inhibit the responding conventional T cells with which they share a partial overlap. Since the conventional T cells can proliferate on binding to viral antigen without local Treg-mediated suppression, larger populations of partially overlapping Tregs are required to maintain a global net proliferation rate of zero. But this process naturally breaks the emergent tiling needed for robustness, and thus could produce increased sensitivity to fluctuations in self-antigen levels. In the future, it will be interesting to further explore this scenario to better understand if it can increase our understanding of autoimmune diseases.

\matmethods{ Please see SI Appendix  for detailed Materials and Methods. Briefly, we constructed a detailed, biologically realistic model for the immune dynamics of Tregs, conventional T cells,  and interleukins. Our model relied of experimentally supported biological assumptions that we outline in detail in the SI. We then derived the minimal model in Eq. \ref{eq:Tc} and Eq. \ref{eq:Treg} presented in the main text  by assuming interleukin dynamics were fast compared to T cell and Treg proliferation rates and that interleukin consumption of Tregs is much larger than those of conventional T cells.  By exploiting the ecological interpretation of this minimal model, we  derived a dual representation of this model in terms of optimization (Eq. \ref{Eq:w_opt}). We then used Eq. \ref{Eq:w_opt} to derive sufficient conditions for emergent tiling using a series of duality transformations. To check these analytic results, we performed numerical simulations of both the dynamical models and corresponding optimization problems. All code in the accompanying github repository \url{https://github.com/Emergent-Behaviors-in-Biology/immune-svm}.}

\showmatmethods{} 

\acknow{We are extremely grateful to Arup Chakraborty for many useful conversations and suggestions on this work, and to Rustom Antia for advice and encouragement in the early phases of the project. This work was funded by a Simons Investigator in MMLS award and NIH NIGMS R35GM119461 grant to PM.}

\showacknow{} 

\bibliography{references}

\end{document}


\title{SI Appendix }
\author{Robert Marsland III}
\author{Owen Howell}
\author{Andreas Mayer}
\author{Pankaj Mehta}
\date{\today}

\maketitle

\renewcommand{\theequation}{S\arabic{equation}}
\renewcommand{\thetable}{S\arabic{table}}
\renewcommand{\thefigure}{S\arabic{figure}}
\setcounter{equation}{0}  
\setcounter{figure}{0}
\setcounter{table}{0}

\section{Details of our mathematical model}
We begin by considering a more mechanistic, biologically plausible model of Treg-mediated adaptive immunity. As in the main text, we will always use the convention that T cells to refer to non-regulatory T cells (i.e. conventional T cells). The basic elements of our model are as follows:

{\color{black} \subsection{Assumptions and basic dynamics of the model}
\begin{itemize}
	\item $\lambda_i$ are clone sizes of ``conventional'' T cells (specifically, CD25$^-$CD4$^+$ helper T cells).
	\item $w_\alpha$ are clone sizes of Tregs.
	\item $v_x $ is the abundance of antigen-presenting cells (APC's) displaying antigen $x$. In general, $x$ can represent any antigen displayed by a class II MHC, including neoantigens, but in this work we focus on the scenario where all displayed antigens are self-peptides.
	\item ${\rm IL}_x$ is the average local concentration of interleukin 2 (IL-2) in the vicinity of APC's displaying antigen $x$. 
	\item $p_{ix}^c$ (``cross-reactivity function'') is the probability that a conventional T cell from clone $i$ that encounters an APC displaying antigen $x$ will bind and activate. The cross-reactivity function is determined by the binding affinity $\Delta G_{ix}$ between the T cell receptor and the antigen, via some model of the binding and activation kinetics. For example, a simple two-state equilibrium model would give $p_{ix}^c = 1/(1+e^{- \Delta G_{ix}/k_B T})$. In the present work, we do not attempt to relate $p_{ix}^c$ to $\Delta G_{ix}$, but instead sample $p_{ix}^c$ directly from one of three probability distributions described in Section IV below.
	\item $p_{ \alpha x}^r$ is the probability that a Treg from clone $\alpha$ that encounters an APC displaying antigen $x$ will bind and activate. See above for explanation of relationship to binding affinity.
	\item IL-2 stimulates proliferation of both Tregs and normal T cells, with the growth rate some saturating functions $g_r({\rm IL}_x),\,\, g_c({\rm IL}_x)$ of the local IL-2 concentration ${\rm IL}_x$.
	\item T cells deplete IL-2 at a rate proportional to the level of growth stimulation, with constants of proportionality $\epsilon_c^{-1}$ and $\epsilon_r^{-1}$ (notation comes from analogy with the efficiency of resource conversion into biomass).
	\item Activated T cells produce IL-2 at rate $a$.
	\item Activated Tregs directly suppress growth of nearby activated T cells, with each Treg cell decreasing the growth rate of T cells in its vicinity by an amount $b$. 
	\item Tregs only suppress conventional T cells bound to the same APC, as suggested by the experiments of \cite{takahashi1998immunologic}. 
	\item Each antigen $x$ is displayed on a small fraction $f \ll 1$ of the total population of APC's. This implies that a Treg binding to antigen $x$ suppresses a conventional T cell bound to a different antigen $y$ only on a much smaller fraction $f^2\ll f$ of APC's. Under these conditions, we can neglect cross-antigen suppression, and consider only the suppression that occurs between Tregs and conventional T cells that are activated by the same antigen $x$.
	\item Both Tregs and conventional T cells circulate rapidly through the body, so that the total populations (including both activated and unactivated cells) are evenly distributed over all APC's.
	\item In the absence of extracellular IL-2, activated T cells proliferate at a basal rate $\rho$. 
	\item Extracellular IL-2 is degraded by some external mechanisms, and has a lifetime $\tau$ in the absence of T cells.
\end{itemize}
These statements result in the following set of differential equations:
\begin{align}
\frac{d\lambda_i}{dt} &= \lambda_i \sum_x v_x p_{ix}^c \left[\rho + g_c({\rm IL}_x)-b \sum_{\alpha} w_\alpha p_{ \alpha x}^r\right]\\
\frac{dw_\alpha}{dt} &= w_\alpha \sum_x v_x p_{ \alpha x}^r\left[ g_r({\rm IL}_x) - m\right]\\
\frac{d{\rm IL}_x}{dt} &= a\sum_i \lambda_i p_{ix}^c - \epsilon_c^{-1} \sum_i \lambda_i p_{ix}^c g_c({\rm IL}_x) - \epsilon_r^{-1} \sum_{\alpha} w_\alpha p_{ \alpha x}^rg_r({\rm IL}_x) - \tau^{-1} {\rm IL}_x.
\end{align}
}
\subsection{Considering limit of fast interleukin dynamics yields minimal model in main text}
To derive the minimal model in the main text, we assume that interleukin dynamics is fast compared to Treg and T cell proliferation. In this case, we can make a quasi-adiabatic approximation by setting $d{\rm IL}_x/dt = 0$. We further assume that the responses of the T cells and Tregs to interleukin concentrations are far from saturation so that we can approximate the growth rates using linear functions $g_c({\rm IL}_x) = c_c {\rm IL}_x, g_r({\rm IL}_x) = c_r {\rm IL}_x$. With these two assumptions, one gets that steady-state concentration of interleukins near antigen $x$ takes the form
\begin{align}
{\rm IL}_x = \frac{a\sum_i \lambda_i p_{ix}^c}{\tau^{-1} + \epsilon_c^{-1} c_c \sum_i \lambda_i p_{ix}^c + \epsilon_r^{-1} c_r \sum_{\alpha} w_\alpha p_{ \alpha x}^r}.
\end{align}
We will work in the regime where Tregs are the main sink for IL-2, and dominate the denominator of this expression. This results in the following dynamics for the two populations:
\begin{align}
\frac{d\lambda_i}{dt} &= \lambda_i \sum_x v_x p_{ix}^c \left[\rho+ \frac{c_c\epsilon_r a}{c_r \sum_\alpha w_\alpha  p_{ \alpha x}^r} \sum_j \lambda_j p_{jx}-b \sum_{\alpha} w_\alpha p_{ \alpha x}^r\right]\\
\frac{dw_\alpha}{dt} &= w_\alpha \sum_x v_x p_{ \alpha x}^r \left[ \frac{\epsilon_r  a}{\sum_\beta w_\beta  p_{ \beta x}^r} \sum_j \lambda_j p_{jx}^c - m\right].
\end{align}
If we additionally assume that  $c_c/c_r$ small (consistent with the assumption of Treg-dominated interleukin consumption), we can ignore the positive feedback term in the first equation, yielding the system of equations:
\begin{align}
\frac{d\lambda_i}{dt} &= \lambda_i \sum_x v_x p_{ix}^c \left[\rho-b \sum_{\alpha} w_\alpha p_{ \alpha x}^r\right]\nonumber\\
\frac{dw_\alpha}{dt} &= w_\alpha \sum_x v_x p_{ \alpha x}^r \left[ \frac{\epsilon_r  a}{\sum_\beta w_\beta  p_{ \beta x}^r} \sum_j \lambda_j p_{jx}^c - m\right],\label{Eq:dynamics0}
\end{align}
which are identical to the dynamics in the main text (where for notational simplicity we write $\epsilon_r a$ simply as $a$).

\begin{figure}
	\includegraphics[width=17cm]{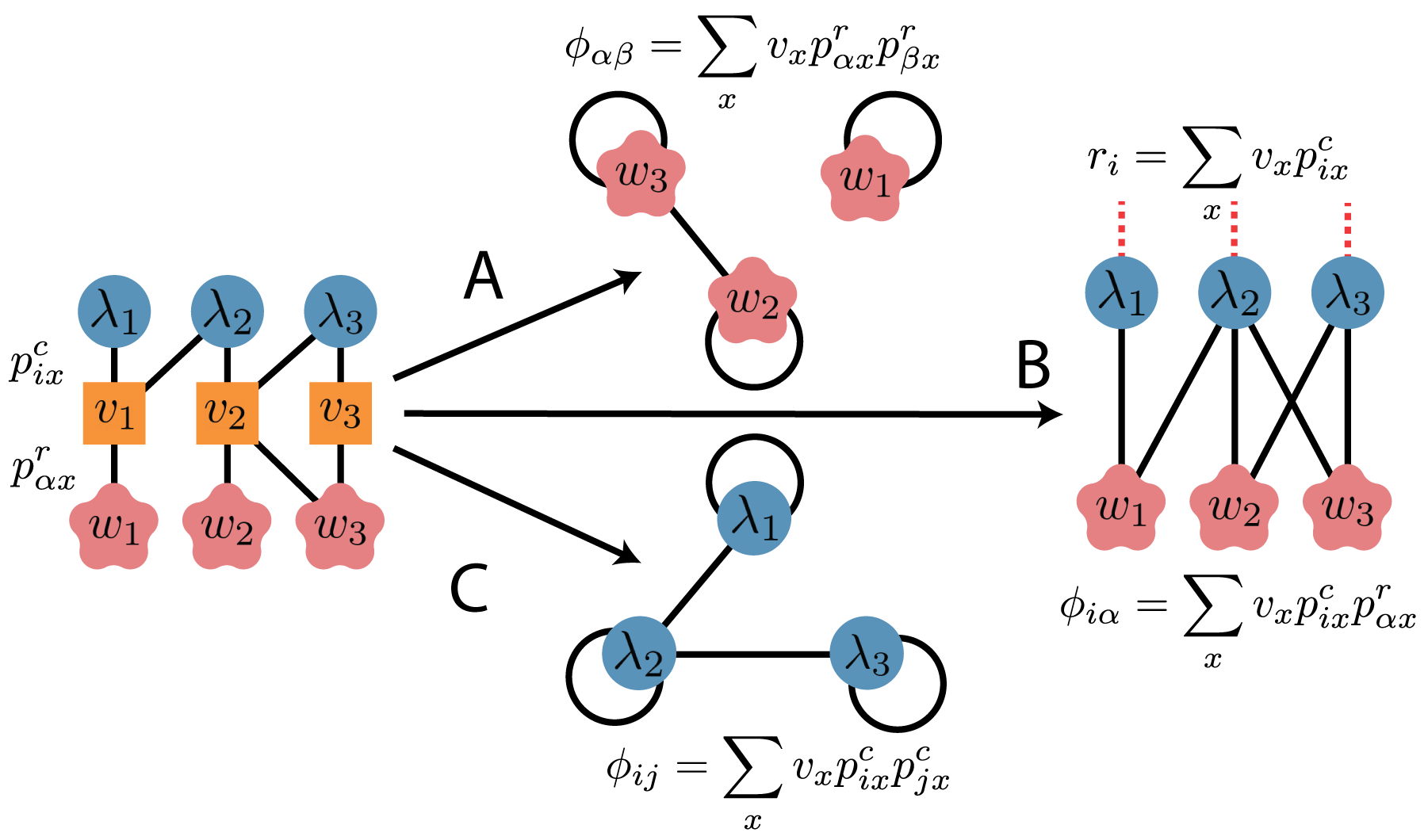}
	\caption{{\bf Defining the overlaps.} The cross-reactivity functions $p_{\alpha x}^r$ and $p_{ix}^c$ define a network of interactions, with edges connecting Tregs and conventional T cells to antigens that can bind to their TCR, and edge weights representing the affinity of the interaction. The strength of the indirect interaction between two T cells can be quantified in terms of the product of their affinities for the same antigen, summed over all antigens and weighted by the antigen abundance. This procedure gives rise to three ``overlap kernels'': (A) $\phi_{\alpha \beta}$ for (competitive) effective interactions between Tregs, (B) $\phi_{ij}$ for (mutualistic) effective interactions between conventional T cells, and (C) $\phi_{i \alpha}$ for effective interactions between conventional T cells and Tregs. Note that $\phi_{ij}$ only appears in the positive feedback term of the full dynamical model defined in the first section of the SI. This term may be important at later stages of the immune response, when conventional T cell populations become large, but is neglected in the present analysis of the initial proliferation dynamics.}
	\label{fig:overlap}
\end{figure}

\subsection{Rewriting our dynamics in terms of overlap kernels}

We now describe the approximation mentioned in the main text that is required to rewrite the above dynamics in terms of overlap kernels. As stated in the main text, the overlap kernels are defined by (see Fig. \ref{fig:overlap}):
\begin{align}
\phi_{\alpha \beta} &= \sum_x v_x p_{ \alpha x}^rp_{ \beta x}^r \nonumber \\
\phi_{i \alpha} &= \sum_x v_x p_{ \alpha x}^rp_{i x}^c \nonumber \\
r_i  &= \sum_x p_{i x}^c v_x.
\label{eq:kernel_defs}
\end{align}

\begin{figure}
	\includegraphics[width=17cm]{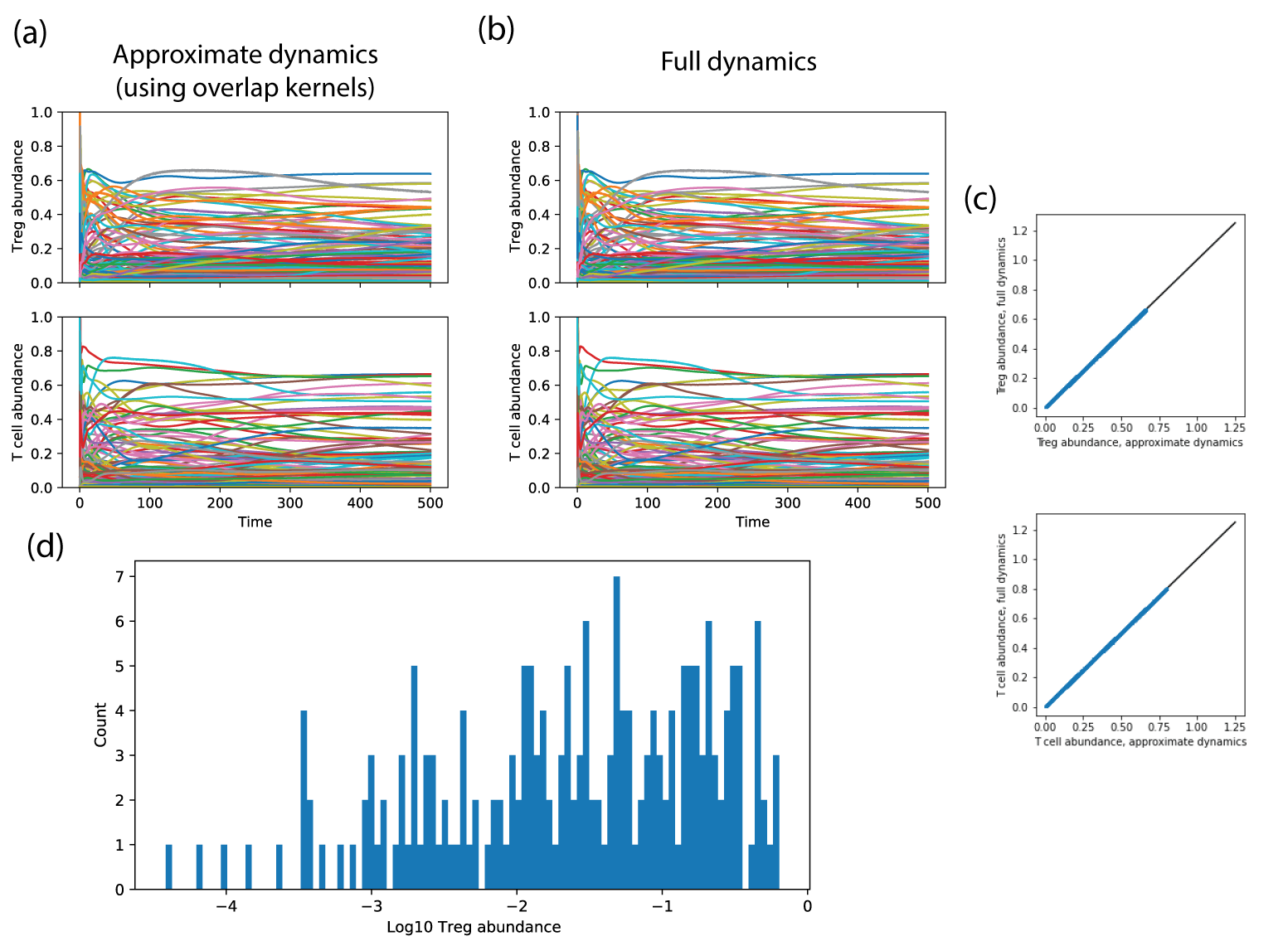}
	\caption{\color{black}{\bf Comparing full dynamics and overlap kernel approximation.} \emph{(a)} Sample trajectories of Treg abundances $w_\alpha$ and conventional T cell abundances $\lambda_i$ using the approximate dynamics of Eq. \ref{Eq:dynamics1}. Cross-reactivity functions were generated using the one-dimensional shape space described in the final section of the SI (``Details on numerical simulations''), with cross-reactivity width $\sigma = 8$, and the other parameters set to $N_a = 100, N_c = 100, N_r = 50, \rho=a=b=m=1$. Initial abundances of Tregs and conventional T cells were sampled from a lognormal distribution with logarithmic mean 0 and logarithmic standard deviation $\sigma = 2$. \emph{(b)} Trajectories of the full dynamics of Eq. \ref{Eq:dynamics0}, using the same cross-reactivity functions, parameter values and initial conditions as the previous panel. \emph{(c)} Comparison of exact and approximate dynamics from the previous two panels. Each point represents the abundance of a single Treg or conventional T cell lineage at a single time point in the two simulations. \emph{(d)} Histogram of final Treg abundances in the simulation of the full dynamics from panel b. Note that the horizontal axis is the base-10 logarithm of the abundance.}
	\label{fig:compare}
\end{figure}

Rearranging Eq. \ref{Eq:dynamics0} yields a set of equations where the cross-reactivity function and antigen concentrations almost always appear within an overlap expression:
\begin{align}
\frac{d\lambda_i}{dt} &= \lambda_i \left(\sum_y v_y p_{iy}^c\right) \left[\rho-b \frac{\sum_{\alpha,x} v_x p_{ \alpha x}^r p_{ix}^c w_\alpha}{\sum_y v_y p_{iy}^c }\right]\\
\frac{dw_\alpha}{dt} &= w_\alpha \left[\epsilon_r  a \sum_{j,x} \frac{v_x p_{\alpha x}^r p_{jx}^c \lambda_j}{\sum_\gamma w_\gamma  p_{ \gamma x}^r}  -m\sum_{\beta,x} \frac{ v_x p_{\alpha x}^r p_{\beta x}^r w_\beta}{\sum_\gamma w_\gamma  p_{ \gamma x}^r} \right].
\end{align}
The final step requires an uncontrolled approximation, whereby we ignore the correlations between the numerators and denominators in the dynamics of $w_\alpha$, and sum over $x$ separately for both sides of the fraction. This approximation is strictly justified only in the emergent tiling regime, where the denominator $\sum_{\gamma} p_{\gamma x} w_\gamma$ is the same (equal to $\rho/b$) for all $x$. But in numerical simulations it appears to work well even outside of this regime, as well as during the transient on the way to an emergent tiling fixed point (see Fig. \ref{fig:compare}).

Using this approximation along with the overlap definitions provided above, we obtain the dynamics stated in Eq. 4 of the main text:
\begin{align}
\frac{d\lambda_i}{dt} &= \lambda_i r_i \left[\rho  -b r_i^{-1} \sum_{\alpha} \phi_{i\alpha} w_\alpha \right] \nonumber \\
\frac{dw_\alpha}{dt} &= \frac{w_\alpha}{\sum_\beta w_\beta \bar{p}_\beta} \left[ \epsilon_r a \sum_j  \phi_{j\alpha} \lambda_j - m \sum_\beta  \phi_{\alpha\beta}w_\beta \right]
\label{Eq:dynamics1}
\end{align}
where $\bar{p}_\beta \equiv \sum_x p_{\beta x}^r$.

\section{Treg dynamics as optimization}

We now show the dynamics above have a natural interpretation in terms of constrained optimization. To do so, we make use of the duality between the steady states of the equations above and  constrained optimization. For completeness, we briefly explain this duality here. Please see our earlier papers \cite{mehta2019constrained,marsland2020minimum}  for a detailed discussion. 

Notice that the steady states of Eq. \ref{Eq:dynamics} satisfy the following equations
\begin{align}
0= \lambda_i  \left[\rho  -b r_i^{-1} \sum_{\alpha} \phi_{i\alpha} w_\alpha \right] \nonumber \\
0=  \left[ \epsilon_r a \sum_j \lambda_j \phi_{j\alpha} - m \sum_\beta w_\beta \phi_{\alpha\beta}\right].
\end{align}
Let us define $\tilde{\lambda}_i={\epsilon_r a r_i \over m b} \lambda_i$. Then we can rewrite the equation above as
\begin{align}
0= \tilde{\lambda}_i  \left[\rho  -b r_i^{-1} \sum_{\alpha} \phi_{i\alpha} w_\alpha \right] \nonumber \\
0=  \left[ -\sum_j \tilde{\lambda}_j b r_j^{-1}  \phi_{j\alpha}-  \sum_\beta w_\beta \phi_{\alpha\beta}\right].
\label{Eq:dynamics}
\end{align}
If we define the functions 
\begin{align}
g_i(\{w_\alpha\}) = \left[\rho  -b r_i^{-1} \sum_{\alpha} \phi_{i\alpha} w_\alpha \right]
\end{align}
and 
\begin{align}
f(\{w_\alpha\})= {1 \over 2} \sum_\beta w_\beta \phi_{\alpha\beta} w_\alpha,
\end{align}
then the steady-state equations take the form
\begin{align}
0 &=  \sum_j \tilde{\lambda}_j  {\partial  g_j (\{w_\alpha\}) \over \partial w_\alpha} - {\partial f(\{w_\alpha\}) \over \partial w_\alpha} \nonumber \\
0 &= \tilde{\lambda}_j g_j (\{w_\alpha\}),
\end{align}

We recognize the equations above as precisely the Karush-Kuhn-Tucker (KKT) conditions for constrained optimization with $f(\{w_\alpha\})$ the function being optimized and the functions $g_j (\{w_\alpha\})$ specifying the constraints. 
Thus, the steady-states of the equations above coincide with the solutions of the following constrained optimization problem:
\begin{align}
\underset{\mathbf{w}}{\rm argmin} \, &\frac{1}{2}\sum_{\alpha,\beta} w_\alpha \phi_{\alpha\beta} w_\beta \nonumber \\
&{\rm subject}\,{\rm to:}\nonumber\\
&r_i^{-1} \sum_\alpha \phi_{i\alpha} w_\alpha \geq \frac{\rho}{b}. \label{eq:bound} \\
& w_{\alpha} \geq 0
\label{Eq:w_opt}
\end{align}

As an aside, this is very nearly single-class SVM, with training data $r_i^{-1} \phi_{i\alpha}$. It finds a hyperplane that separates all the data from the origin while maximizing the distance between the plane and the origin. It would be exactly a single-class SVM if $\phi_{\alpha\beta}$ were the identity matrix and the $w_{\alpha} \geq 0$ condition was not enforced. The $w_{\alpha} \geq 0$ slightly changes the geometrical interpretation of the 1-class SVM. Specifically, the requirement that $w_{\alpha} \geq 0$ forces the simplex $  \sum_{\alpha} w_{\alpha} \phi_{\alpha} - p $ to have all positive coordinate intercepts. See \cite{howell2020machine} for more details on this interpretation.

\section{Emergent tiling as solution to optimization problem}

In the previous section, we stated the optimization problem in terms of the overlap kernels $\phi_{i \alpha}$ and $\phi_{\alpha \beta}$, which integrate over the whole antigen space. The emergent tiling conditions in the main text, however, involve the antigen space explicitly (Eq. 6). In this section, we exploit dual formulations of the optimization problem to highlight the role of the antigen space and make the connection to the emergent tiling conditions more transparent. 

Before proceeding,  it will be useful to collect some of the basic definitions that we stated above through out the text:
Denote the dimension of the T-cell space $N_c$ and the dimension of the Treg space $N_r$. Furthermore, denote the ``naive" dimension of antigen space by $N_a$. This is essentially the number of $x$ we sum over.   However, if the matrix $p_{ix}^c$  and $p_{ \alpha x}^r$ are structured so that there are lots of correlations between the different antigens $x$,  than this naive counting might be quite misleading, and we should really thing about the effective dimension of the antigen space $N_a^{\rm eff}$. If all the $x$ are uncorrelated, then of course $N_a^{\rm eff} =N_a$.

Also, in this section we will drop the tilde from $\tilde{\lambda}_i$ (defined in the previous section), with the understanding that the solutions obtained for $\lambda_i$ must be multiplied by $mb/(\epsilon_r a r_i)$ in order to give the actual T cell populations.

We will now write the optimization in Eq. \ref{Eq:w_opt} in a slightly different way, and in the process gain more physical insights about what we mean by these conditions. We begin by noting that plugging in the first line of Eq. \ref{eq:kernel_defs} and flipping signs results in the trivial rewriting
\begin{align}
\underset{\mathbf{w}}{\rm argmax} \, &-\frac{1}{2} \sum_x  v_x \left( \sum_{\alpha} p_{ \alpha x}^r w_\alpha \right)^2 \nonumber \\
&{\rm subject}\,{\rm to:}\nonumber \\
&r_i^{-1} \sum_\alpha \phi_{i\alpha} w_\alpha \geq \frac{\rho}{b}. \nonumber \\
& w_{\alpha} \geq 0.
\label{Eq:w_opt2}
\end{align}

\subsection{Reformulation 1}
It is now straightforward to check that 
\begin{align}
\underset{\mathbf{w}}{\rm argmax} \, &-\frac{1}{2} \sum_x  v_x \left( \sum_{\alpha} p_{ \alpha x}^r w_\alpha \right)^2 
\end{align}
is the same as the following max-min optimization
\begin{align}
\underset{\{ s_x\} }{\rm argmin} \hspace{0.1in}\underset{\mathbf{w}}{\rm argmax} \, & \sum_x  {s_x^2 \over 2 v_x}   -  \sum_{\alpha, x }s_x  p_{ \alpha x}^r w_\alpha 
\label{Eq: sw_opt}
\end{align}
To see this note that we can differentiate this with respect to $s_x$ and set this expression to zero to get that the optimum over the new auxiliary variable is
\begin{align}
s_x^{\rm opt}= \sum_{\alpha} p_{ \alpha x}^r w_\alpha v_x,
\label{Eq:def_sx_opt}
\end{align}
and plugging this into Eq. \ref{Eq: sw_opt} gives the original optimization problem Eq. \ref{Eq:w_opt2}. Thus, we see that $s_x^{opt}$ just measures the total coverage of antigen $x$
by Tregs. 

Another useful manipulation is to note that 
\begin{align}
\sum_\alpha \phi_{i\alpha} w_\alpha= \sum_{x}  p_{i x}^c \sum_{\alpha} v_x p_{ \alpha x}^r w_\alpha  = \sum_x s_x^{\rm opt} p_{ix}^c
\end{align}
Combining this with the last line of  Eq. \ref{Eq:w_opt2} we can rewrite the constraint $r_i^{-1} \sum_\alpha \phi_{i\alpha} w_\alpha \geq \frac{\rho}{b}$ as
\begin{align}
\sum_x p_{i x}^c v_x \left( \sum_\alpha p_{ \alpha x}^r w_\alpha -\frac{\rho}{b} \right) =\sum_x p_{ix}^c \left( s_x^{\rm opt} - v_x  \frac{\rho}{b} \right) \ge 0
\end{align}
Notice that one way of satisfying this constraint is by requiring $s_x^{\rm opt}/v_x = \sum_{\alpha} p_{ \alpha x}^r w_\alpha = \frac{\rho}{b}$. Here we see that the idea is that we will make sure that
each site $x$ gets the same amount of coverage, set by $\rho/b$. This is precisely the emergent tiling we are seeking. 

This might not always be possible since in general the dimension of antigen space, $N_a$ may be larger than the dimension of the T-cell space $N_c$ and the dimension of the Treg space $N_r$. However, if the matrix $p_{ \alpha x}^r$ is structured such that the number of Tregs is much larger than the effective dimension of antigen space $N_a^{\rm eff}$ then this can be inverted. More generally, the existence of a solution is governed by a Gardner like transition analogous to that of perceptrons \cite{gardner1988space,Nishimori,landmann2018systems}.

\subsection{Reformulation 2: Langrange multipliers instead of inequality constraints}
In this section, we will reformulate the problem again and get even more insight into how we can view the problem in the antigen space. This will also lead to important clues about where some unexpected properties of this solution come from. Let us start with Eq. \ref{Eq:w_opt_Lagrange} and rewrite it in terms of $p_{ix}^c$, $p_{ \alpha x}^r$, and $v_x$ using expressions in Eq. \ref{eq:kernel_defs}:
\begin{align}
\underset{\{\lambda_i\}}{\rm argmax} \hspace{0.1in} \underset{\mathbf{w}}{\rm argmin} \, &   \sum_x  \frac{v_x}{2} \left(\sum_{\alpha} p_{ \alpha x}^r w_\alpha \right)^2  - \sum_x v_x \left( \sum_i \lambda_i p_{i x}^c\right)\left( \sum_\alpha p_{x \alpha} w_\alpha \right)  + \sum_i \lambda_i \ r_i \frac{\rho}{b} \\
&{\rm subject}\,{\rm to:} \, \lambda_i \geq 0, \, w_\alpha \ge 0
\label{Eq:w_opt_Lagrange}
\end{align}
Let us focus on the quantity we are optimizing. Notice that by completing the square and changing sign we can rewrite this as 
\begin{align}
 \frac{v_x}{2} \left[ \left(\sum_{\alpha} p_{ \alpha x}^r w_\alpha \right)- \left( \sum_i \lambda_i p_{i x}^c\right) \right]^2 - \frac{v_x}{2} \left( \sum_i \lambda_i p_{i x}^c\right)^2
\end{align}
Let us introduce two new auxiliary variables $A_x$ and $B_x$ that will couple to each of these square terms. Then notice the expression above can be written as
\begin{align}
&\frac{v_x}{2} \left[ \left(\sum_{\alpha} p_{ \alpha x}^r w_\alpha \right)- \left( \sum_i \lambda_i p_{i x}^c\right) \right]^2 - \frac{v_x}{2} \left( \sum_i \lambda_i p_{i x}^c\right)^2 \nonumber \\
=&\, \underset{\{A_x\}}{\rm argmax} \,  {-A_x^2 \over 2 v_x}- A_x \left[ \left(\sum_{\alpha} p_{ \alpha x}^r w_\alpha \right)- \left( \sum_i \lambda_i p_{i x}^c\right) \right]- \frac{v_x}{2} \left( \sum_i \lambda_i p_{i x}^c\right)^2 \nonumber \\
&\underset{\{A_x\}}{\rm argmax} \,\underset{\{B_x\}}{\rm argmin}\,   {-A_x^2 \over 2 v_x}- A_x \left( \sum_{\alpha} p_{ \alpha x}^r w_\alpha- \sum_i \lambda_i p_{i x}^c \right) +  {B_x^2 \over 2 v_x}- B_x  \sum_i \lambda_i p_{i x}^c
\end{align}

With all these manipulations we can rewrite the original optimization problem as 
\begin{align}
\underset{\{A_x\}}{\rm argmax} \, \,\underset{\{B_x\}}{\rm argmin}\, \,\underset{\{\lambda_i\}}{\rm argmax} \,\, \underset{\{w_\alpha\}}{\rm argmin} \, &
\sum_x \left[  {-A_x^2 \over 2 v_x}- A_x  \left( \sum_{\alpha} p_{ \alpha x}^r w_\alpha - \sum_i \lambda_i p_{i x}^c \right)+  {B_x^2 \over 2 v_x}- B_x  \sum_i \lambda_i p_{i x}^c \right] +  \sum_i \lambda_i \ r_i \frac{\rho}{b} \nonumber \\
&{\rm subject}\,{\rm to:} \, \lambda_i \geq 0, \, w_\alpha \ge 0
\label{Eq:fulldecouple_opt}
\end{align}
It doesn't look like we have done much right now. But one nice thing about this new optimization function is that it is \emph{linear} in the $w_\alpha$ and $\lambda_i$. We can then take derivatives with respect to all four quantities to get a set of optimization equations. Taking the derivative with respect to $B_x$ yields
\begin{align}
B_x^{\rm opt} = \sum_i v_x p_{ix}^c \lambda_i^{\rm opt}.
\label{Eq:opt-Bx}
\end{align}
In other words, $B_x^{\rm opt}$ just measures the coverage of antigen site $x$ by all conventional T cells. Taking the derivative with respect to $A_x$ gives
\begin{align}
A_x^{\rm opt} &= \sum_i v_x p_{ix}^c \lambda_i^{\rm opt}- \sum_\alpha v_x p_{x \alpha} w_\alpha ^{\rm opt} \nonumber \\
&= B_x^{\rm opt}- \sum_\alpha v_x p_{x \alpha} w_\alpha ^{\rm opt}
\label{Eq:opt-Ax}
\end{align}
This equation say that $A_x^{\rm opt}$ is just the difference between the Tcell and Treg coverages at antigen site $x$. Taking the derivative with respect
to $w_\alpha$ yields
\begin{align}
\sum_x p_{x \alpha} A_x^{\rm opt}=0.
\label{Eq:opt-omega1}
\end{align}
To understand the meaning of this equation, it is useful to combine this with Eq. \ref{Eq:opt-Ax} and use Eq. \ref{eq:kernel_defs} to rewrite this as 
\begin{align}
\sum_i \phi_{i \alpha} \lambda_i^{\rm opt} - \sum_\beta \phi_{\alpha \beta} w_\beta^{\rm opt}=0,
\label{Eq:opt-omega2}
\end{align}
which is simply the statement that the growth rate of Treg $\alpha$ must be zero.
Finally, taking the derivative with respect to $\lambda_i$ gives the equation
\begin{align}
\sum_x (B_x^{\rm opt} - A_x) p_{ix}^c = r_{i} {\rho \over b}.
\label{Eq:opt-lambda1}
\end{align}
 Plugging in Eq. \ref{Eq:opt-Ax} and using Eq. \ref{eq:kernel_defs} it is easy to see that this equation just states that the Tcell growth rates should be zero.\\\\

 \subsection{Ansatz for solution to optimization problem}
 Thus, far we haven't gained much. But we will focus on what particularly interesting set of potential solutions to this problem. If the number of antigens $N_a$ is very large compared to the number of T cells $N_i$, then  we can in general  easily find solutions to   $\sum_x B_x^{\rm opt} p_{ix}^c = r_{i} {\rho \over b}$. However, we will make an even stronger ansatz.
 Notice that $r_i= \sum_x p_{ix}^c v_x$ so that the following ansatz is a solution to Eq. \ref{Eq:opt-lambda1}
 \begin{align}
 B_x^{\rm opt} &= v_x {\rho \over b} \nonumber \\
 A_x &=0.
 \label{Eq:ansatz}
\end{align}
In order for these to be good solutions, from Eq. \ref{Eq:opt-Ax} and   Eq. \ref{Eq:opt-Bx} we must have that there exist solutions for $w_\alpha$ and $\lambda_i$ satisfying the following set of equations.
 \begin{align}
 \sum_{\alpha} p_{ \alpha x}^r w_\alpha^{\rm opt} &= {\rho \over b} \nonumber \\
 \sum_{j} p_{j x} \lambda_j^{\rm opt} &= {\rho \over b}.
 \end{align}

In general, these equations may not be solvable since the naive number of antigens $N_a$ could be larger than the number of T cells,  $N_c$, or number of Tregs $N_r$. But as long as the ``effective'' dimensionality of the antigen space (accounting for correlations between antigen binding affinities) satisfies $N_a^{eff} \ll N_c, N_r$, then we should be able to find such a solution. In fact, such a criteria has recently been derived in the statistical physics literature \cite{landmann2018systems}. When the cross-reactivities are i.i.d, in the thermodynamic limit where $N_a, N_r, N_c \gg 1$, there is a phase transition between a regime where such a solution exists and does not depending on the ratios of $N_r/N_a$ and $N_c/N_a$. This phase transition corresponds exactly to the Gardner solution to the perceptron problem \cite{gardner1988space,Nishimori,landmann2018systems}.

\subsection{Relation to Gardner Transition in Perceptrons}
{\color{black}
Here, we give a brief overview of the Gardner transition in perceptrons and discuss the relationship to the problem at hand (see books by Nishimore and Engel for details about perceptrons and statistical learning) \cite{Nishimori,engel2001statistical}. The perceptron problem is concerned with ``storing'' $P$ boolean patterns $\xi^u$, with each pattern $\mu$ consisting of $N$ input bits $S_i^\mu = \pm 1$ for $1 \le i\le N$. Each pattern is assigned a value $R^\mu=\pm 1$ according to the rule $R^\mu = \mathrm{sgn}(\sum_i J_iS_i^\mu)$. A natural question one can ask is what is the maximum number of patterns such a function can classify correctly, where we are allowed to choose the $N$ parameters $J_i$,

It was found that if the patterns were chosen randomly, that the maximum number of patterns that any function of this form could classify correctly was exactly equal to $P=2N$. For $P>2N$ generically, there existed no choice of $J$ that will classify all $P$ patterns correctly. This ``phase transition'' marks a boundary to the regime where there are no solutions for the $N$ variables $J_i$ to the $P$ equations 
\begin{align}
R^\mu = \mathrm{sgn} (\sum_i J_i \epsilon_i^\mu).
\end{align}
If $P<2N$, there exists a set of $J_i$ that solve these equations. On the other hands, if $P>2N$, there exists no solutions to this problem 

As can be clearly seen, this system of equations are isomorphic to the kinds considered in our immunological problems with $N$ playing the role of the Treg dimension, $P$ the antigen dimension, and $w_\alpha^{\rm opt}$ playing the roles of $J_i$. Thus, if the Treg dimension is at least twice the antigen dimension, we will have solutions to the equations for emergent tiling derived above. Since technically, we need to solve two such set of equations, we also need the diversity of conventional Tcells to be at least twice that of the antigen dimension. However, we make the implicit, biologically realistic assumption that the conventional Tcell diversity is always larger than the Treg diversity.  This correspondence between perceptrons and ecology was also noted in  a replica calculation in \cite{landmann2018systems}.
}

\subsection{Insensitivity to antigen concentrations}
\begin{figure}
	\includegraphics[width=17cm]{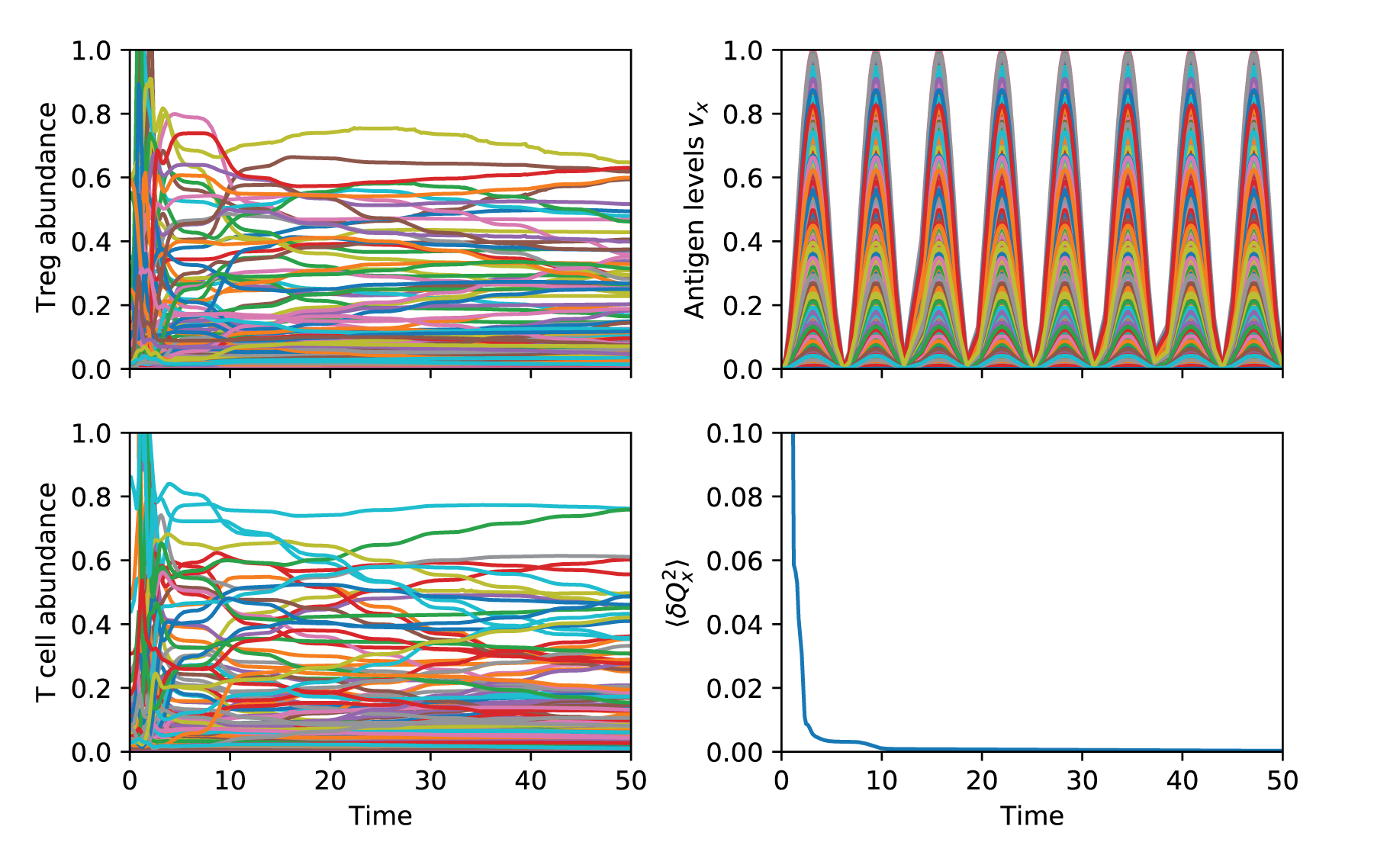}
	\caption{\color{black}{\bf Simulation with rapidly oscillating $v_x$.} The full dynamics of Eq. (\ref{Eq:dynamics0}) were integrated with the antigen abundances $v_x$ varying in time as $v_x = v_x^0 \sin^2(\omega t)$. The oscillation amplitudes $v_x^0$ were sampled from a uniform distribution between 0 and 1, and the oscillation frequency was $\omega = 0.5$. The other parameters and initial conditions sampling were identical to those of Figure \ref{fig:compare}b above. Abundances of conventional T cells and Tregs are shown as a function of time, along with the time-varying antigen abundances $v_x$ and the Treg coverage nonuniformity $\langle \delta Q_x^2\rangle$.}
	\label{fig:vx}
\end{figure}

Such a solution if it exists has some amazing properties, namely the T cell and Treg growth rates are insensitive  to the antigen concentrations $v_x$ ensuring that the Tregs exhibit an emergent tiling over the T cells:
\begin{itemize}
 \item It balances conventional T cell and Treg activity at every antigen site $x$ independently since $A_x=0$.
 \item The growth rates of the T cells and Tregs become insensitive to changes in the $v_x$. 
 \end{itemize}
 \vspace{0.1in}
 To see this latter point, notice that we can define ``growth rates'' for T cells $g_i$ and Tregs $g_\alpha$ as 
\begin{align}
\frac{d\lambda_i}{dt} &\equiv \lambda_i g_i ={1 \over b} \lambda_i \left[ r_i\rho  - \sum_{\alpha} \phi_{i\alpha} w_\alpha \right]\label{eq:lam}\\
\frac{dw_\alpha}{dt} &\equiv \frac{w_\alpha}{\sum_\beta w_\beta \bar{p}_\beta} g_\alpha = \frac{w_\alpha b}{\rho} \left[ \epsilon_r a \sum_j \lambda_j \phi_{j\alpha} - m \sum_\beta w_\beta \phi_{\alpha\beta}\right].
\end{align}
A straightforward calculations using the definitions above then yields:
\begin{align}
{\partial g_i \over \partial v_x} &=  \sum_x p_{ix}^c \left[ {v_x \rho \over b}+A_x-B_x \right]=0 \nonumber \\
{\partial g_\alpha \over \partial v_x} &= \sum_x p_{ \alpha x}^rA_x =0
\end{align}

\section{Details on Numerical Simulations}
\begin{figure}
	\includegraphics[width=17cm]{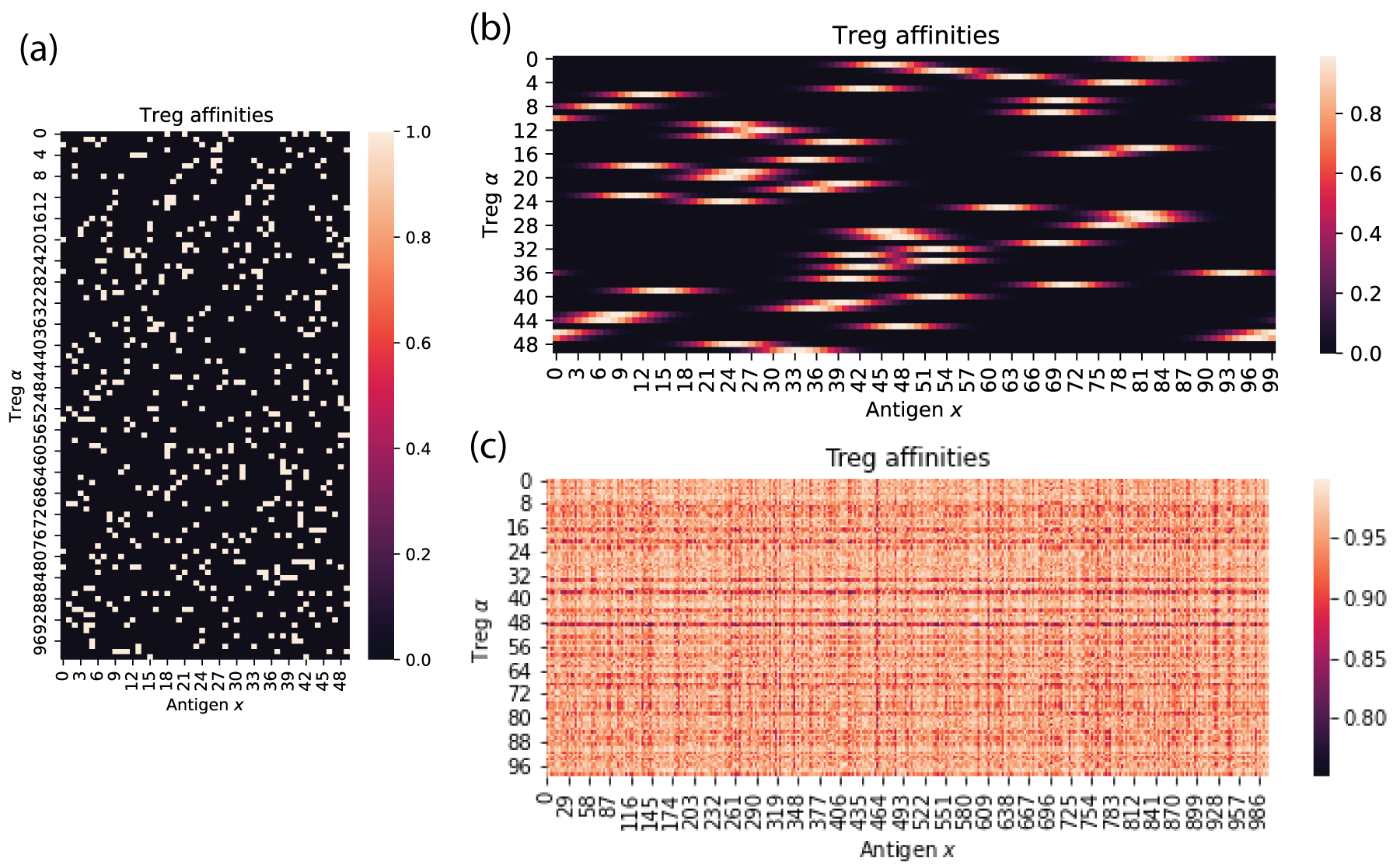}
	\caption{\color{black} {\bf Random sampling of cross-reactivity functions $p_{ix}^c$ and $p_{\alpha x}^r$.} Representative samples from the three cross-reactivity function ensembles described in SI Section IV are shown as heat maps. In all simulations, cross-reactivity functions for conventional T cells and Tregs were drawn from the same distribution. \emph{(a)} Bernoulli distribution, in which a randomly chosen T cell and antigen have a probability $p$ (here equal to 0.1) of interacting, and all interactions are assigned independently. \emph{(b)} One-dimensional shape space, where a given T cell can bind to antigens whose shape coordinate is within a tolerance $\sigma$ (here equal to 4) of a randomly assigned optimal shape. \emph{(c)} High-dimensional shape space, where antigens and TCR's are randomly assigned coordinates in a shape space of specified dimension (here equal to 5), and the binding probability is determined by the pairwise distances.}
	\label{fig:cross-reactive}
\end{figure}

To generate Figure 3  of the main text, we generated random cross-reactivity functions according to two different protocols, and then used the optimization formulation of the equilibrium conditions (Eqs. 5 of the main text) to efficiently obtain equilibrium populations of Tregs and conventional T cells for each realization. Scripts for generating the matrices, solving the optimization problem and plotting the results can be found in the accompanying github repository \url{https://github.com/Emergent-Behaviors-in-Biology/immune-svm}. Optimization was performed using the Python package CVXPY \cite{cvxpy}.

In the first protocol, the elements of $p_{\alpha x}^r$ and $p_{ix}^c$ were sampled from Bernoulli distributions, with success probability equal to 0.1, 0.2 or 0.3. 30 realizations were generated for 20 values of $N_r$ ranging from 100 to 300. The other parameters were $N_a = 100, N_c = 1,000, \rho=m=a=b=c_r=v_x=1$.

In the second protocol, the elements of $p_{\alpha x}^r$ and $p_{ix}^c$ were chosen in a correlated way, encoding a one-dimensional ``shape space.'' Specifically, we generated a Gaussian cross-reactivity shape centered at the midpoint of the shape space, given by
\begin{align}
p_x &= e^{-(x-N_a/2)^2/2\sigma^2}
\end{align}
for a given cross-reactivity width $\sigma$, and then shifted it by a random offset $x_\alpha$ for each Treg and $x_i$ for each conventional T cell. The shifts were performed with periodic boundary conditions, so that all Tregs and T cells still had the same overall binding capacity. 10 realizations were generated for each of 100 values of $\sigma$, ranging from 10 to 100. For each sampled cross-reactivity matrix $p_{\alpha x}^r$, we defined an effective number of distinguishable antigens $N_a^{\rm eff}$ by counting the number of singular values above an empirically determined numerical cutoff threshold of $\epsilon = 10^{-6}$. The horizontal axis in the right-hand panels of Fig. 3 is given by $N_r/N_a^{\rm eff}$. The true number of antigens was $N_a = 5,000$, and the other parameters were $N_c = 500, \rho=m=a=b=c_r=v_x=1$.

\begin{figure}
	\includegraphics[width=17cm]{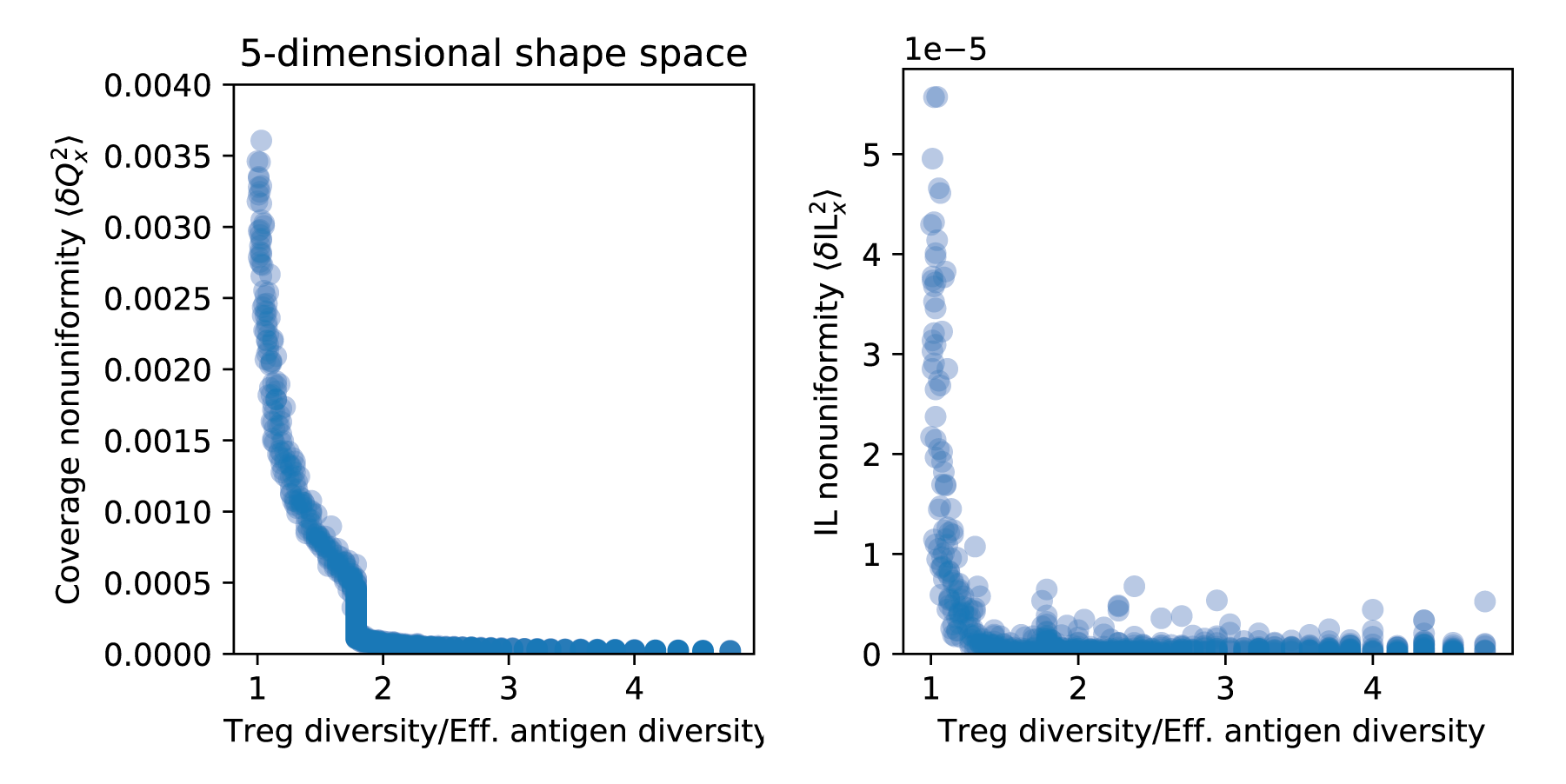}
	\caption{\color{black}{\bf Emergent tiling transition in high-dimensional shape space.} Same as Figure 3 of the main text, but with the cross-reactivity functions $p_{\alpha x}^r$ and $p_{ix}^c$ sampled using a five-dimensional shape space as illustrated in Figure \ref{fig:cross-reactive} above and described in SI Section IV. The effective antigen diversity was defined as the number of singular values of $p_{\alpha x}$ exceeding a cutoff threshold of $\epsilon = 10^{-5}$. }
	\label{fig:highshape}
\end{figure}

We also ran simulations with cross-reactivity functions generated in a higher-dimensional shape space. Following \cite{mayer2015well}, we assigned shape coordinates $\mathbf{a}_x$ and $\mathbf{r}_i$ (or $\mathbf{r}_\alpha$ for Tregs) in a space of dimension $D$ to the antigens and TCR's, respectively. We then calculated the cross-reactivity functions as
\begin{align}
p_{ix}^c &= e^{-\frac{||\mathbf{r}_i-\mathbf{a}_x||^2}{2\sigma^2}}\\
p_{\alpha x}^r &= e^{-\frac{||\mathbf{r}_\alpha-\mathbf{a}_x||^2}{2\sigma^2}}
\end{align}
where a small distance in shape space corresponds to a good fit between TCR and antigen, while larger distances produce bad fits that do not bind. The parameter $\sigma$ sets the radius of the region of shape space that is compatible with a given TCR. We randomly sampled all the shape coordinates ($\mathbf{r}_i$, $\mathbf{r}_\alpha$ and $\mathbf{a}_x$) from unit normal distributions. For the simulations shown in Figure \ref{fig:highshape}, we sampled $N_r = 100$ Tregs, $N_c = 100$ conventional T cells and $N_a = 1000$ antigens. We chose a shape space of dimension $D = 5$, following dimensionality estimates derived from analysis of hemagglutination inhibition assays for influenza \cite{lapedes2001geometry}. We varied the cross-reactivity width $\sigma$ from 5 to 20 in order to sweep the ratio of the Treg diversity to effective antigen diversity through the emergent tiling threshold.

\bibliography{references}